\documentclass[reqno,11pt]{amsart}
\usepackage{amsmath, latexsym, amsfonts, amssymb, amsthm, amscd}
\usepackage{sidecap}
\usepackage{graphics,epsf,psfrag} 
\numberwithin{equation}{section}

\newtheorem{theorem}{Theorem}[section]
\newtheorem{lemma}[theorem]{Lemma}

\newtheorem{rem}[theorem]{Remark}

\newcommand{\R}{\mathbb{R}}

\renewcommand{\tilde}{\widetilde}

\newcommand{\cF}{{\ensuremath{\mathcal F}} }

\newcommand{\cL}{{\ensuremath{\mathcal L}} }

\newcommand{\cU}{{\ensuremath{\mathcal U}} }
\newcommand{\cV}{{\ensuremath{\mathcal V}} }
\newcommand{\cR}{{\ensuremath{\mathcal R}} }

\DeclareMathSymbol{\leqslant}{\mathalpha}{AMSa}{"36} % nicer `smaller or equal'
\DeclareMathSymbol{\geqslant}{\mathalpha}{AMSa}{"3E} % nicer `larger or equal'
\DeclareMathSymbol{\eset}{\mathalpha}{AMSb}{"3F}     % nicer `emptyset'
\renewcommand{\leq}{\;\leqslant\;}                   % redef. of < or =
\renewcommand{\geq}{\;\geqslant\;}                   % redef. of > or =
\newcommand{\dd}{\,\text{\rm d}}             % a straight d for differentials
\newcommand{\bbN}{{\ensuremath{\mathbb N}} }

\newcommand{\bbR}{{\ensuremath{\mathbb R}} }
\newcommand{\bbS}{{\ensuremath{\mathbb S}} }

\newcommand{\bbZ}{{\ensuremath{\mathbb Z}} }

\newcommand{\ga}{\alpha}

\newcommand{\gd}{\delta}
\newcommand{\gep}{\varepsilon}       % \ge already exists...

\newcommand{\gG}{\Gamma}

\newcommand{\gD}{\Delta}

\newcommand{\gl}{\lambda}

\newcommand{\gs}{\sigma}

\makeatletter
\def\captionfont@{\footnotesize}
\def\captionheadfont@{\scshape}

\long\def\@makecaption#1#2{%
  \vspace{2mm}
  \setbox\@tempboxa\vbox{\color@setgroup
    \advance\hsize-6pc\noindent
    \captionfont@\captionheadfont@#1\@xp\@ifnotempty\@xp
        {\@cdr#2\@nil}{.\captionfont@\upshape\enspace#2}%
    \unskip\kern-6pc\par
    \global\setbox\@ne\lastbox\color@endgroup}%
  \ifhbox\@ne % the normal case
    \setbox\@ne\hbox{\unhbox\@ne\unskip\unskip\unpenalty\unkern}%
  \fi
  \ifdim\wd\@tempboxa=\z@ % this means caption will fit on one line
    \setbox\@ne\hbox to\columnwidth{\hss\kern-6pc\box\@ne\hss}%
  \else % tempboxa contained more than one line
    \setbox\@ne\vbox{\unvbox\@tempboxa\parskip\z@skip
        \noindent\unhbox\@ne\advance\hsize-6pc\par}%
\fi
  \ifnum\@tempcnta<64 % if the float IS a figure...
    \addvspace\abovecaptionskip
    \moveright 3pc\box\@ne
  \else % if the float IS NOT a figure...
    \moveright 3pc\box\@ne
    \nobreak
    \vskip\belowcaptionskip
  \fi
\relax
}
\makeatother
\def\writefig#1 #2 #3 {\rlap{\kern #1 truecm
\raise #2 truecm \hbox{#3}}}

\begin{document}
\title[Transitions in  active rotator systems]{
Transitions in  active rotator systems:\\
invariant hyperbolic manifold  approach}

\author{Giambattista Giacomin}
\address{
  Universit{\'e} Paris Diderot (Paris 7) and Laboratoire de Probabilit{\'e}s et Mod\`eles Al\'eatoires (CNRS),
U.F.R.                Math\'ematiques, Case 7012 (site Chevaleret)
             75205 Paris Cedex 13, France
}
%\email{giacomin\@@math.jussieu.fr}
\author{Khashayar Pakdaman}
\address{
  Institut Jacques Monod, Universit\'e Paris 7--Denis Diderot,
Bat. Buffon, 15 rue H\'el\'ene Brion - 75013 Paris, France
}

\author{Xavier Pellegrin}
\address{Institut Jacques Monod, Universit\'e Paris 7--Denis Diderot,
Bat. Buffon, 15 rue H\'el\'ene Brion - 75013 Paris, France}

\author{Christophe Poquet}
\address{
  Universit{\'e} Paris Diderot (Paris 7) and Laboratoire de Probabilit{\'e}s et Mod\`eles Al\'eatoires (CNRS),
U.F.R.                Math\'ematiques, Case 7012 (site Chevaleret)
             75205 Paris Cedex 13, France
}

\date{\today}

\begin{abstract}
Our main focus is on a general class of
   active rotators with mean field interactions, that is  globally coupled large families of
 dynamical systems on the unit circle with non-trivial stochastic dynamics. The dynamics of each isolated 
 system 
  is $\dd \psi_t =-\gd V'(\psi_t)\dd t + \dd w_t$, where $V'$ is a periodic function,
  $w$ is a Brownian motion and $\gd$ is an intensity parameter.
 It is well known that 
 the interacting dynamics is accurately described, in the limit of infinitely many interacting components, by a Fokker-Planck PDE
 and the model reduces for $\gd=0$ to a particular case of the Kuramoto synchronization model,
 for which one can show the existence of a stable normally hyperbolic manifold of stationary solutions
 for the corresponding Fokker-Planck equation
 (we are interested in the case in which this manifold is non-trivial, that happens when the interaction 
 is sufficiently strong, that is in the  synchronized regime  of
  the Kuramoto model). We use the robustness of normally hyperbolic structures to
 infer qualitative and quantitative results on the $\vert\gd\vert \le \gd_0$ cases, with
 $\gd_0$ a suitable threshold: as a matter of fact, we obtain  an accurate 
 description of the dynamics on the invariant manifold for  $\gd \neq0$ and
 we link it explicitly to the potential $V$.
  This approach allows to 
 have a complete
 description of the phase diagram of the active rotators model, at least for $\vert \gd\vert \le \gd_0$, 
 thus identifying for which values of the parameters (notably, noise intensity and/or coupling strength)
  the system exhibits periodic pulse waves or stabilizes at a quiescent resting state. Moreover, some of our results are very explicit and this brings a new insight into the combined effect of active rotator dynamics,
  noise and interaction. The links with the literature on  specific systems, notably neuronal models, are discussed in detail.  
   \\
  \\
  2000 \textit{Mathematics Subject Classification: 37N25, 82C26, 82C31,  92B20}
  \\
  \\
  \textit{Keywords: Active rotator model, Coupled excitable systems, Interacting diffusions, Fokker-Planck PDE, Normally hyperbolic manifolds, Pulsating waves, Neuronal models }
\end{abstract}

\maketitle
\section{Introduction}

\subsection{Coupled excitable systems}
There are diverse examples of threshold phenomena in natural systems.
Dynamics of excitable systems, as exemplified by neuronal membranes (to which we restrict
for sake of conciseness),
constitute one of the common forms of threshold behavior. Excitable
systems are characterized by their nonlinear response to perturbations. In
the absence of inputs, they remain at a resting state. This state is
locally stable in the sense that the system returns rapidly to it after
small perturbations. However, for inputs beyond a critical range, the
response of the system takes on a very different form, before regaining
the resting state. In the phase portrait of the system, subthreshold
responses correspond to monotonic returns to the stable equilibrium while
suprathreshold ones appear as excursions that take the system transiently
away from the stable equilibrium. Excitability is one of the key neuronal
properties at the heart of signal processing and transmission in nervous
systems. Motivated by their ubiquity and numerous experimental
observations attesting to their functional importance, there has been a
characterization of  various forms of excitability in terms of the
geometry of the phase portrait of dynamical systems \cite{cf:izhikevich}.

Excitable systems are particularly sensitive to noise because such random
signals contain consecutive sub and suprathreshold segments that occur in
an unpredictable manner. The interplay between the nonlinearity inherent
in the threshold mechanisms and the noise induced fluctuations can produce
a 
large variety %plethora 
of dynamics in excitable systems, some of which are reviewed in
\cite{cf:lindner}. In this paper, we consider one of these, namely, noise
induced synchronous coherent oscillations in assemblies of coupled
excitable systems.

Noisy excitable systems display irregular repetitive suprathreshold
excursions henceforth referred to as firing. In ensembles of such units
receiving independent noise, the firings of the units remain independent
from one another as long as there are no interconnections between them.
Coupling the units with one another introduces correlations between their
firings. Synchrony is the extreme form of such correlations when the units
fire almost simultaneously. However, synchronous firings can be irregular. One of
the surprising effects of noise in assemblies of interacting excitable
systems is that for some range of coupling strength and noise intensity,
units fire synchronously and regularly. 
The wide occurrence of these noise
induced coherent dynamics and their underlying mechanisms are well
documented as explained below. Their putative functional role in nervous
systems is to participate in rhythm generation in the absence of pacemaker
units (see for instance \cite{cf:kosmidis}). Despite the large number of
numerical explorations devoted to this phenomenon, it has not been
analyzed from a mathematical standpoint. The purpose of the present paper
is to deal with this aspect.  

Two key elements are at play in the occurrence of noise induced regular
synchronous firing in assemblies of interacting excitable units, one is
that interacting excitable units act globally like a single excitable
system at the population level, the other is that noise driven excitable
systems undergo coherence resonance \cite{cf:pikovsky}. How the combination of these two
phenomena leads to noise induced regularly synchronous firing has been
first highlighted in an analysis of  networks of an elementary neuronal
model \cite{cf:pham98a,cf:pham98b}, see also \cite{cf:han,cf:rappel}.

The important point is the generality of this mechanisms. It neither
relies on the refined properties of specific classes of excitable systems
nor on the types of coupling. In fact, noisy  assemblies of all common
neuronal models, irrespective of the type of excitability, and whether
coupled  diffusively  or through excitatory pulses or synapses,
readily produce noise induced regular synchronous firing. To our
knowledge, one of the %seldom cited 
earliest reports of this phenomenon
goes back to the explorations of MacGregor and Palasek of randomly
connected populations of neuromimes incorporating a large array of
individual neuronal properties \cite{cf:macgregor}. More recent examples
include the description of the same phenomenon in common neuronal models
such as the Hodgkin-Huxley  \cite{cf:wang}, the FitzHugh-Nagumo
\cite{cf:toral}, the Morris-Lecar  \cite{cf:han}, the Hindmarsch-Rose
\cite{cf:du} and others implementing detailed biophysical properties
\cite{cf:kosmidis}. In these references, besides differences in the models
there are also differences in coupling and network architecture: in some
the units are diffusively coupled, in others they are coupled through
excitatory pulses; in some connectivity is all-to-all, while others deal
with random networks. Our enumeration, which does not
intend to be exhaustive, illustrates the ease with which assemblies of
excitable units generate noise-induced synchronous regular activity,
irrespective of model and network details.

The ubiquity of the phenomenon strongly supports investigating its key
characteristics through the mathematical analysis of a minimal model that
captures its essence. The model we consider is a general version of
the so-called active rotator (AR) which is representative of the so-called 
class {\sl I} excitable systems \cite{cf:izhikevich}. 

\subsection{Active rotator models}
The  AR  is a variant of the Kuramoto
model  for excitable-oscillatory systems that evolve on a unit circle
\cite{cf:acebron}. Precisely, the AR model can be introduced via the stochastic equations
\begin{equation}
\label{eq:Ndiffusion}
\dd \psi_j (t) \,=\, - \gd V' \left( \psi_j (t)\right) \dd t - \frac K N\sum_{i=1}^N
\sin\left( \psi_j(t) - \psi_i(t)\right) \dd t + \gs \dd w_j(t)\, ,
\end{equation}
where $j=1, \ldots, N$, $N$ is a (large) integer, $K$, $\gs$, and $\gd $ are non-negative constants, the $w_j$'s are IID standard Brownian motions and $V$ is a smooth function
(in the applications  the case in which $V'$ is   a trigonometric polynomial
will play an important role, so me may as well think of this case). We look at
$\psi_j$ as an element of $\bbS:= \bbR /2\pi \bbZ$, that is $\psi_j$ is a phase, and, of course we have to supply
an initial condition for \eqref{eq:Ndiffusion}: for example we can take $\{ \psi_j(0)\}_{j=1, \ldots, N}$ to be independent
identically distributed random variables. 

This set of equations defines a diffusion on $\bbS^N$ describing the evolution of $N$ 
noisy interacting phases:
 note that since $K\ge 0$ the interaction has a tendency to  {\sl synchronize} the $\psi_j$'s
and let us stress from now that such an $N$-dimensional diffusion reduces for $\gd=0$
to a dynamics that is reversible with respect to the Gibbs measure with Hamiltonian given by
$-\frac KN \sum_{i, j} \cos(\psi_i-\psi_j)$ and inverse temperature $\gs^{-2}$. Such a Gibbs measure goes under the name 
of ``mean field classical XY  model": we refer to \cite{cf:BGP} for more details, but we point out that for $\gd >0$
(of course the case $\gd<0$ is absolutely analogous), unless $V$ is a periodic 
function  (which we do not assume: consider for example $V'(\psi)=1$), the dynamics is not reversible.  
Nevertheless,  it is  well known 
that the large $N$ behavior of such a system can be described in terms of
the Fokker-Planck or McKean-Vlasov PDE (the literature on this issue is very vast: see for example the
 references in \cite{cf:BGP}):
\begin{equation}
\label{eq:aim}
\partial_t p^\gd_t (\theta) \, =\, \frac {\gs^2}2 \partial_\theta^2 p^\gd_t(\theta) - \partial_\theta \left[ p^\gd_t(\theta) (J * p^\gd_t)(\theta)
\right] + \gd \partial_\theta \left[p^\gd_t(\theta) V'(\theta)\right]\, , 
\end{equation}
where $J(\cdot):= -K \sin(\cdot)$ and $\theta\in \bbS$. To  be precise, $p^\gd_t(\cdot)$ is a probability density
and it captures the $N \to \infty$ limit of the empirical (probability) measure
$\frac 1N \sum_{j=1}^N \gd_{\psi_j(t)}(\dd \theta)$, where $\gd_a$ is the Dirac delta measure on $a$.
Actually,  one can even describe with great accuracy (as $N \to \infty$) the dynamics
of each unit system (in interaction!): it evolves following a non-local diffusion equation, called 
at times {\sl non-linear diffusion}. The non-locality comes from the fact that
 $\psi_j$ is subject not only to the force field $V'$, but also to the field corresponding 
 to the interaction with all other unit systems, and it all boils down to
 \begin{equation}
 \label{eq:nonlMP}
 \dd \psi(t) \, =\, - \gd V' ( \psi(t)) \dd t + (J*p_t^\gd)(\psi(t)) \dd t + \gs \dd w(t)\, ,
 \end{equation} 
 with $w$ a standard Brownian motion, 
and it turns out that the probability distribution of $\psi (t)$ is precisely $p_t^\gd$ if
$\psi(0)$ has distribution $p_0^\gd$.

\medskip

In mathematical terms, the question that we want to tackle is: what is the relation between the 
simple deterministic one dimensional dynamics $\dot \psi= -V' (\psi)$ (Isolated Deterministic System: IDS)
and the behavior of the associated $N$ dimensional diffusion, for $N$ large? 
The question is actually twofold. First, given a potential $V$ for the IDS, 
what is the collective dynamic of the $N$ large limit \eqref{eq:aim}? 
Conversely, what are the possible collective dynamics of \eqref{eq:aim}?
In order to be more concrete let us ask the following sharper questions: is it possible that
\begin{itemize}
\item the IDS  has only one stable point, for example if $V(\psi)= \psi-a \cos(\psi)$ for $a>1$,  but the $N\to \infty$ system exhibits stable 
periodic behavior, that is there is a stable periodic solution to \eqref{eq:aim}?
\item the IDS has only periodic solutions,  but the $N\to \infty$ system has stable
stationary solutions? 
\end{itemize}
\medskip

The fact that the answer to these questions is positive is, to a certain extent, known.
Notably, in their numerical investigations of
the dynamics of coupled noisy ARs, Shinomoto and Kuramoto reported the
existence of collective periodic oscillations, the same phenomenon we have
referred to as noise induced regular synchronous activity
\cite{cf:shinomoto1986a}. They also performed numerical explorations of
the transitions to and from this coherent state. The key ingredient in
such analyses has been to consider the bifurcations of the associated
 Fokker-Planck equation (we anticipate that  our results make rigorous 
 some of their predictions, see Section~\ref{sec:ARs}). 
 %In \cite{cf:shinomoto1986a}, numerical evidence for saddle-node and Hopf bifurcations were presented.
%Numerical bifurcation analyses were further extended in
%\cite{cf:sakaguchi}  who explored the dynamics of the system in wider
%parameter ranges providing evidence for the occurrence of a Takens-Bogdanov bifurcation. 
%These investigations hinted to the existence of noise induced regular synchronous activity in coupled ARs and parameter ranges where it would be observed. 
To clarify how noise generates such
time-periodic global activity in coupled excitable ARs, Kurrer and
Schulten approximated the solutions of the nonlinear Fokker-Planck
equations by Gaussian distributions \cite{cf:kurrer}. Under this
assumption, they obtained closed ordinary differential equations for the
mean and variance of the distribution and used the bifurcation diagram of
these to investigate the regimes where the model generates periodic
oscillations. 
Related work can be found for example in \cite{cf:hasegawa,cf:ohta}, where finite $N$ 
analysis has been performed, or in \cite{cf:park,cf:paullet,cf:shimoto1986b}, where
variants of the model have been considered.

%All these bifurcations concern the infinite size system. Ohta and Sassa have focussed on fluctuations in finite size systems near the saddle node transition \cite{ohta}. Hasegawa has also considered finite size fluctuations \cite{cf:hasegawa}. Coupled ARs are a generic model for investigating the dynamics of populations of excitable systems. There are many other variants of the noisy AR assemblies that are beyond the scope of the present work such as for instance ARs with multiplicative rather than additive noise \cite{cf:park} or populations on lattice rather than all-to-all coupling  \cite{cf:shinomoto1986b}\cite{cf:paullet} that can sustain periodic waves even in the absence of noise.

\medskip

However, from a mathematical viewpoint this phenomenon is only very partially understood. 
We are aware of the contributions \cite{cf:RSV,cf:Sch2,cf:Sch3,cf:touboul} that are somewhat close in spirit to what we are doing:
these references deal with periodic behavior in nonlinear Markov processes and, more generally, with the effect of the noise on (mean field) interacting dynamical systems. We also deal with nonlinear Markov
processes -- the evolution equation \eqref{eq:nonlMP} contains the law of the process itself -- even if this aspect is not emphasized in the remainder of the paper. 
In particular,  Scheutzow \cite{cf:Sch2}  provides examples of mean-field type systems 
in which periodic behavior  arises in  the $N \to \infty$ system, even if it is not present in
absence of noise.
The ingenious model set forth in \cite{cf:Sch2} is however rather particular: for example the author plays with 
some stochastic 
differential equations of 
nonlinear Markov type that admit also  Gaussian solutions and the analysis boils down to studying the behavior of the expectation and covariance of these solutions.  This is close to the approach  taken by Touboul, Hermann and Faugeras 
\cite{cf:touboul}, who extensively exploit the {\sl preservation of the Gaussian character} that holds  for certain nonlinear Markov processes and they do so for models that aim at describing neural activity. We stress that in their approach the 
IDS dynamics is linear, while for us the nonlinearity of the  IDS is a key feature. 
 Rybko, Shlosman and Vladimirov in \cite{cf:RSV}  study 
a connected network of servers that behaves in a periodic fashion in the infinite volume limit, when
there are sufficiently many customers per server ({\sl load per server}): in this regime  there is an effective synchronization between servers and the {\sl load per server} plays a role which is similar to the parameter $K$ in our work, cf. \eqref{eq:Ndiffusion}.

%We signal also \cite{cf:Sch3}, that deals with a non-linear system -- the Brusselator -- but the result in this case is the persistence of the limit cycle present in the  IDS, under coupling and noise.

\subsection{Informal presentation of approach and  results}
The purpose of this work is to show that for general AR systems one can systematically
(at least for some range of the parameters)
and quantitatively exhibit the  relation between the IDS and the infinite system.
This is done by showing that the (infinite-dimensional!) AR system does behave like
a one dimensional AR, and the  latter can be throughly analyzed. 
We obtain such 
 a drastic reduction of dimension
 by exploiting the
  fact that  for the  $\gd=0$ case of \eqref{eq:aim} one can 
perform a rather detailed analysis (due to the fact that it is the grandient
flow of a free energy  functional \cite{cf:BGP}).
In that case and when $K>K_c:=\gs^2$, stationary solutions of (\ref{eq:aim}) are the constant  $\frac{1}{2\pi}$ which is unstable, and 
a circle $M=\{q(\cdot-\theta_0): \, \theta_0 \in \bbS\}$, which is a manifold of non-constant invariant solutions: these solutions describe the synchronized state of the oscillators that have a tendency to be close to $\theta_0$. 
The function $q: \bbS \to (0, \infty)$ is explicitly known and one can show that $M$ is stable.
In fact it has been shown that $M$ is stable in the sense that it is a  {\sl stable normally hyperbolic manifold} for the 
$\gd=0$ evolution
(See Section~\ref{sec:SNHM}).  A deep  result  in dynamical systems theory guarantees the robustness of
normal hyperbolicity under suitable perturbations \cite{cf:HPS}, see also \cite{cf:BLZ,cf:SellYou}: this means that, if $\gd>0$ is not too large,
there exists an invariant manifold $M_\gd$ which is stable and normally hyperbolic for the evolution \eqref{eq:aim},
and $M_\gd$ is a {\sl smooth deformation} of $M$.  
In particular, for small enough $\gd$, $M_\gd$ is still a one dimensional manifold diffeomorphic to a circle, 
and the phase along this manifold plays the role of the natural phase $\psi \in \mathbb{S}$ 
of the IDS $\dot \psi= -V' (\psi)$.  This makes clearly a direct link between the 
(one dimensional) IDS 
and the $N=\infty$ system (\ref{eq:Ndiffusion}), which  is an infinite dimensional dynamical system.

\begin{figure}[hlt]
\begin{center}
\leavevmode
\epsfxsize =14 cm
\psfragscanon
\psfrag{p}{$p(\cdot)$}
\psfrag{q}{$\!q_\psi(\cdot)$}
\psfrag{phi}{$\psi$}
\psfrag{th}{$\theta$}
\psfrag{pi}{$\pi$}
\psfrag{-pi}{$-\pi$}
\psfrag{0}{$0$}
\epsfbox{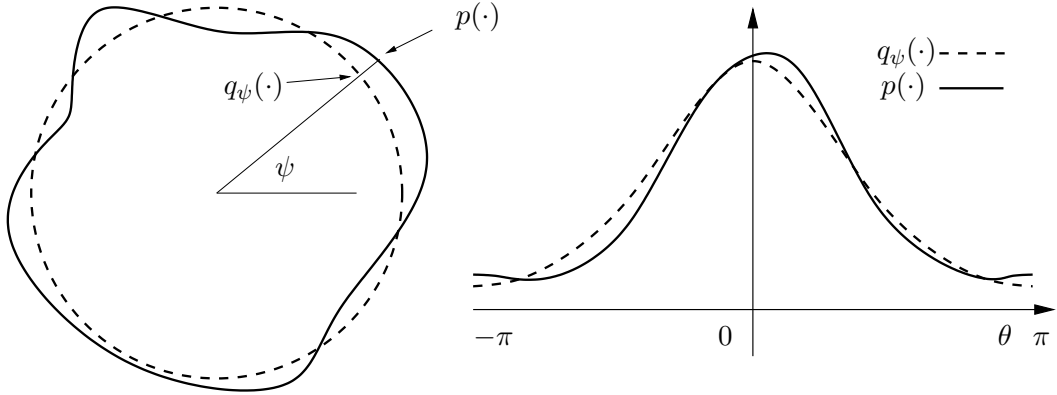}
\end{center}
\caption{For $\gd$ sufficiently small the solutions of \eqref{eq:aim} with an initial condition
in a $L^2$-neighborhood of $M$ (in the figure on the left
$M$ is drawn by a dashed line), that is an initial condition close to a
$q_\psi(\cdot)=q(\cdot+ \psi)$, stay close to $M$  for all times. In fact, they are attracted
by a manifold $M_\gd$ (solid line, still on the left) that is a small (and smooth) deformation of $M$. For every function $q_\psi(\cdot)$
in $M$ one associates only one function $p(\cdot)$ on $M_\gd$. While the image on the right
stresses the function viewpoint, the one on the left stresses the geometric viewpoint: $M$ is a circle
and it is hence parametrized just by one parameter (the phase $\psi$), but $M_\gd$ can also
be reduced simply to $\psi$. The dynamics on the two manifolds is hence  reduced to
the dynamics of $\psi$, with the substantial difference that even if the dynamics for
$\gd=0$ is trivial in the sense that $M$ is a manifold of stationary solutions
of \eqref{eq:aim} with $\gd=0$, for $\gd>0$ the dynamics on $M_\gd$ can be non-trivial. As a matter
of fact, we are going to show that by playing on the choice of $V(\cdot)$ {\sl essentially any}
phase dynamics can be observed  on $M_\gd$, and this 
for every $K$ and $\gs $ such that $K>\gs^2$.}
\label{fig:phase}
\end{figure}

The type of results that we obtain is well exemplified in the 
most basic of the active rotator models, namely the one in which we take  
 $V(\psi)=\psi -a \cos (\psi)$ (without loss of generality: $a\ge 0$): note that, for $a <1$, the IDS describes just a rotation on the circle, while for $a> 1$ the IDS has a stable point ($\psi=-\arcsin\left(\frac{1}{a}\right)$), to which it is driven, unless
sitting on the unstable stationary point $\psi =\arcsin\left(\frac{1}{a} \right)+\pi$.  
Let us keep in mind that $M_\gd$ is close to $M$, which is a circle, so that also the dynamics on $M_\gd$ 
can be reduced to the dynamics of a phase (see Fig.~\ref{fig:phase}). 
We are going to show in particular that
\begin{enumerate}
\item there exists (in fact, we give it explicitly)  $a_0>1$ 
 such that for $a\in (1,a_0)$ (so the IDS has a stable stationary point!) there exist $K_-,K_+>1$, with $K_-<K_+$ 
such that for $K\in (K_-,K_+)$, and $\gd>0$ sufficiently small \eqref{eq:aim}
 has a stable periodic solution -- a {\sl pulsating wave} -- which corresponds to the fact that the dynamics
 on $M_\gd$ is periodic. For $K \in (1, K_-)$ or $K>K_+$ instead the dynamics on the manifold $M_\gd$ has (only) one
 stable stationary point, so \eqref{eq:aim} has a stable stationary solution (like the IDS).
\item for every $a \in (0,1)$, that is the IDS is rotating, one can find $K_0>1$ (sufficiently close to $1$) such that
whenever $K \in (1, K_0)$ for $\gd$ sufficiently small the dynamics on  $M_\gd$ has (only) one
 stable stationary point. 
\end{enumerate}

\medskip

Actually, these examples are just instances of  results that we will establish for general
 potentials $V$. For example we will show that
  for any $V$ such that $V'$ changes sign (so the IDS has a stable point), for $K$ large enough, the dynamics on the invariant curve stabilizes at an equilibrium for small $\delta$. Or that for any $V$ 
  such that $V'>0$
  (so the IDS is rotating), but 
  with nonzero first harmonic coefficient(s), for $K$ close to $1$ the dynamics on the invariant curve stabilizes at an equilibrium for small $\delta$.

 Finally, regarding the inverse problem, that is the range of possible dynamics, we show that given a noise and a coupling strength such that the $\gd=0$ system exhibits synchronization,
{\sl any} (phase) dynamics can be produced on $M_\gd$, for $\gd$ sufficiently small, by a suitable choice of the IDS dynamics 
(that is, of $V$) and the relation between these two dynamics is explicit.

\section{Mathematical set-up and main results}
\subsection{On the reversible Kuramoto PDE}

Let us first sum up a number of results about
\begin{equation}
\label{eq:Kuram}
\partial_t p^0_t (\theta) \, =\, \frac 12 \partial_\theta^2 p^0_t(\theta) - \partial_\theta \left[ p^0_t(\theta) (J * p^0_t)(\theta)
\right] \, , 
\end{equation}
where  $J(\theta):= -K \sin(\theta)$. Note that we have set $\gs=1$: there is of course no loss of generality
in doing this.
We start by  introducing
the weighted $H_{-1}$ spaces that  are going to play an important role in the sequel. 

Given a positive smooth function $w: \bbS \to (0, \infty)$ we define 
the Hilbert space $H_{-1,w}$ as the closure of the set of smooth functions
 $\bbS \to \R$ such that $\int_\bbS u =0$ with respect to the squared norm
 $\Vert u \Vert_{-1,w}^2:=\int_\bbS w \cU ^2$, where $\cU= \cU_w$ is the primitive of $u$ such that
 $\int_\bbS w\cU =0$. The alternative way to introduce such a space is in terms of {\sl rigged 
  Hilbert spaces} %\cite{cf:Brezis} 
  can be found in \cite{cf:BGP}.
  When $w(\cdot)\equiv 1$ we simply write $H_{-1}$. Let us remark immediately 
  that 
  \begin{equation}
  \label{eq:equiv}
  \Vert u \Vert_{-1, w_1}^2\, =\,\int_\bbS w_1 \left( \cU_{w_2} -\frac{\int_\bbS w_1 \cU_{w_2}}{\int _\bbS w_1}  
  \right) ^2 \, \le \, \int_\bbS w_1 \cU_{w_2}^2 \, \le \, \left\Vert \frac{w_1}{w_2}\right\Vert_\infty \Vert u\Vert_{-1,w_2}^2\, ,
   \end{equation}
so that the the norms we have introduced are all equivalent. 
We will also use the affine space
\begin{equation}
\tilde H_{-1}:=\left\{\frac{1}{2\pi}+u: \, u\in H_{-1}\right\}\, ,
\end{equation}
provided
with the $H_{-1}$ distance. The companion space $\tilde H_1$, defined in the analogous way,
will also appear later on.

\subsubsection{Basic features and the stationary solutions of the reversible Kuramoto PDE}
The reversible Kuramoto PDE has a number of features that we recall here. First of all, the reversible Kuramoto PDE has strong regularizing 
 properties \cite{cf:GPP}, so that we can safely talk about classical smooth solutions for all positive times,
 for example whenever the initial condition is in $L^2$. In particular, 
  \eqref{eq:Kuram} defines an $L^2$-semigroup. Actually, the conservative character of the dynamics
  and the fact that we are dealing with probability distributions naturally lead to work on the affine space
  \begin{equation}
  \label{eq:L2_1}
  L^2_1\, :=\, \{f\in L^2:\,  \int f =1\}\, ,
  \end{equation}
with the $L^2$ distance.
   One of 
 the main feature of \eqref{eq:Kuram}, directly inherited from being the limit
 of a reversible stochastic dynamics, is that it is the gradient flow of a  free energy
 (which is therefore a Lyapunov functional for the evolution).
These properties underlie what follows but we do not directly use them, and so we refer to \cite{cf:BGP},
see also \cite{cf:GPP} for related results.

What plays a direct role in our analysis is the fact that 
all the stationary solutions of \eqref{eq:Kuram} can be written as 
\begin{equation}
\frac 1Z\, {\exp(2Kr \cos(\cdot-\psi))}\, ,
\end{equation}
 where $\psi\in \bbS$ (this accounts for the rotation 
invariance of \eqref{eq:Kuram}), $Z$ is the normalization constant
(fixed by the requirement of  working with probability densities) and $r\ge 0$ is a solution
of the fixed point problem
\begin{equation}
\label{eq:rfixed}
r \, =\, \Psi (2Kr)\, \phantom{move} \text{with } \ \ 
\Psi(x)\, :=\, \frac{\int_\bbS \cos(\theta) \exp(x \cos (\theta))\dd \theta }{\int_\bbS  \exp(x \cos (\theta))\dd \theta } \, .
\end{equation}
$\Psi(0)=0$, so that $r=0$ is a solution of the fixed point problem and $\frac 1{2\pi}$ is a stationary solution. Moreover
$\Psi(\cdot)$ is increasing and concave on the positive semi-axis, so that there exists at most one positive fixed point $r$
and such a fixed point exists if and only if $K>K_c=1$ (see \cite{cf:BGP} and references 
therein). So for $K>1$ (that we assume from now on) there is a manifold, in fact a curve, of stationary solution, besides the constant solution:
\begin{equation}
 M\, :=\, \left\{ q_{\psi}(\cdot)= q_0(\cdot-\psi): \, \psi \in \bbS \right\}\ \ \text{ with } 
\ q_0 (\theta) \, :=\, \frac{\exp\left( 2Kr \cos (\theta)\right)}{\int _{\bbS} \exp\left( 2Kr \cos (\theta)\right) \dd \theta}\, ,
\end{equation}
where $r=r(K)$ is the positive fixed point of \eqref{eq:rfixed}.
We will come back to the manifold structure of $ M$, but we point out that 
the main result in \cite{cf:BGP} means that $M$ is a {\sl stable normally hyperbolic} manifold 
(\cite[p.~494]{cf:SellYou}: we are going to detail this just below): 
we stress that $M$ is actually a manifold of stationary solutions and not only an invariant manifold. 
The key point is that if $p^0_t=q \in M$ the linearized evolution operator 
\begin{equation}
-L_q u \, :=\, \frac 12 u'' - [ u J*q + qJ*u]'\, ,
\end{equation}
with domain $\{u\in C^2(\bbS , \R):\, \int_\bbS u=0\}$ is symmetric in $H_{-1, 1/q}$
and  its closure, that we still call $L_q$, is a self-adjoint operator  operator with compact resolvent,
hence the spectrum is discrete. 
Actually the spectrum is in $[0, \infty)$: $L_q q'=0$ and $q'$ generates the whole kernel of $L_q$: 
the spectral gap is therefore positive and it will be denoted $\gl_K$
(see \cite{cf:BGP} for a proof of all these facts and for an explicit lower bound on $\gl_K$).

In such a framework it is useful to take advantage of some of the interpolation spaces associated 
to $L_q$. For us the (Hilbert) spaces $V_q$  and $V_q^2$ with norms
\begin{equation}
\label{eq:Vq}
\Vert v \Vert_{V_q}\, :=\, 
\left\Vert \sqrt{1+ L_q} \, v\right\Vert_{-1,1/q}\, \ \ \text{ and } \ \
\Vert v \Vert_{V^2_q}\, :=\, 
\left\Vert (1+ L_q) \, v\right\Vert_{-1,1/q}\, ,
\end{equation} 
will play an important role.
In \cite{cf:BGP} it is shown that if $v\in L^2_0:= \{v\in L^2:\, \int_\bbS v=0\}$
\begin{equation}
\label{eq:comp}
c_K \Vert v\Vert_2 \, \le \, \Vert v \Vert_{V_q} \, \le \, 
c_K^{-1} \Vert v\Vert_2
 \, ,
\end{equation}
where here (and below) 
$c_K$ denotes a suitable positive constant that depends only on $K$
(it will not keep the same value through the text: in particular, in this case it is the same for every 
 $q\in  M$).
%Therefore $V_q$ and $L^2_0$, as sets,  coincide. 
Note that if $v\in \cR(L_q)$, $\cR(\cdot)$  denotes the range of $\cdot$,
the spectral gap guarantees that 
\begin{equation}
\label{eq:fspg}
\Vert v \Vert_{V_q}^2 \, \le \, \left(1 + \frac 1{\gl_K} \right) 
\left \Vert \sqrt {L_q} \, v \right\Vert_{-1,1/q}^2 \, .
\end{equation}
At this point it is also worth observing also that, by \eqref{eq:equiv}, 
there exists $c_K>0$ such that for every $\psi_1, \psi_2 \in \bbS$ we have
\begin{equation}
\label{eq:compq}
c_K \Vert v \Vert_{-1,1/q_{\psi_1}}
\, \le \, \Vert v \Vert_{-1,1/q_{\psi_2}}\, \le c_K^{-1}\Vert v \Vert_{-1,1/q_{\psi_1}}\, .
\end{equation}
Of course, we have the analogous estimates in the case in which $1/q_{\psi_2}$, or
$1/q_{\psi_1}$, is replaced by $1$.

\subsubsection{Stable normally hyperbolic manifolds}
\label{sec:SNHM}
We now  quickly review the notion of 
 {\sl stable normally hyperbolic  manifold}, in the $L^2_1$ set-up,  because it will play a central role in our results.
 For this we need a dynamics:
 what we have in mind is \eqref{eq:main} but for the moment 
 let us just think of an evolution semigroup in $L^2_1$
 that gives rise to $\{ u_t\}_{t\ge 0}$, with $u_0=u$,
  to which we can 
 associate a linear evolution semigroup $\{\Phi(u, t)\}_{t \ge 0}$ in $L^2_0$, satisfying  
 $\partial_t  \Phi(u, t)v =A(t)  \Phi(u, t)v$ and $\Phi(u, 0)v=v$, where $A(t)$ is the operator
 obtained by linearizing the evolution around $u_t$.

 For us a stable normally hyperbolic  manifold $M\subset L^2_1$ (in reality we are interested only
 in $1$-dimensional manifolds, that is curves, but at this stage this does not really play a role)
 of characteristics $\gl_1$, $\gl_2$ ($0\le \gl_1< \gl_2$) and $C>0$ is  
 a $C^1$ compact connected manifold which is invariant under the dynamics 
 and for every $u \in M$ there exists a projection $P^o(u)$ on the 
 tangent space of $M$ at $u$, that is $\cR(P^o(u))=:T_uM$, which, for $v \in L^2_0$, satisfies the properties
 \begin{enumerate}
 \item  for every  $t\ge 0$ we have 
 \begin{equation}
 \Phi(u,t) P^o (u_0)v\, =\, P^o (u_t)\Phi(u,t) v\, ,
 \end{equation}
 \item we have  
 \begin{equation}
 \label{eq:shyp1}
 \Vert \Phi(u,t) P^o(u_0) v\Vert_2 \, \le \, C \exp( \gl_1 t)   
 \Vert v \Vert_2\, ,
 \end{equation}
 and, for 
 $P^s:=1-P^o$, we have 
 \begin{equation}
 \label{eq:shyp2}
 \Vert \Phi(u,t) P^s(u_0) v\Vert_2 \, \le \, C \exp( -\gl_2 t)   
 \Vert v \Vert_2\, ,
 \end{equation}
 for every $t\ge 0$;
 \item 
 there exists a negative continuation of 
 the dynamics  $\{ u_t\}_{t \le 0}$ and of the linearized
 semigroup
 $\{ \Phi(u, t) P^o(u_0) v \}_{t \le 0}$ and for any such continuation
 we have
 \begin{equation}
 \label{eq:shyp3}
 \Vert \Phi(u,t) P^o(u_0) v\Vert_2 \, \le \, C \exp( -\gl_1 t)   
 \Vert v \Vert_2\, ,
 \end{equation}
 for $t \le 0$.
 \end{enumerate}

\medskip

As an example -- for us  a crucial example --
let us show that $M$ is a stable normally hyperbolic manifold 
for the $L^2_1$-semigroup  associated to \eqref{eq:Kuram} (in Section~\ref{sec:full} 
we give some details on this semigroup in the general case).
As we have seen, $M$ is an invariant manifold: it is in fact a set of stationary solutions,
so that the dynamics has a (trivial) negative continuation, and it is easy to provide an 
explicit atlas, compatible with the $L^2$ topology,  for which $M$ is a
$C^\infty$ manifold and $T_{q}M_0= \{ a q':\, a \in \bbR \}= \cR(P^o_q)$.
The projection $P^o$ we choose is defined  by 
\begin{equation}
P^o (q)v =P^o_qv\, :=\, 
\frac{
(v, q')_{-1,1/q} \, q'}{(q',q')_{-1,1/q} }\, ,
\end{equation} 
and, since $L_q q'=0$ for every $q \in M$, we see that 
$\gl_1$ can be chosen equal to zero and any value $C\ge 1$ will do. 
Moreover if we set
$v_t:= \Phi (q,t) P^s_q v \in \cR(P^s_q)$ then
\begin{multline}
\Vert v_t \Vert _2 \, \le \, c_K \Vert v_t \Vert_{V_{q}}
\, \le \, c_K \sqrt{1+1/\gl_K} \left\Vert \sqrt{L_q} \, v_t \right\Vert_{-1,1/q}
\\
 \le \, c_K  \sqrt{1+1/\gl_K} \exp\left(-\gl_K t\right)
\Vert\sqrt{L_q}\,  v_0 \Vert_{-1,1/q} \, \le \, c'_K \exp\left(-\gl_K t\right) \Vert v\Vert_2\, ,
\end{multline}
 where we have used \eqref{eq:compq}, then \eqref{eq:fspg}, then 
 the spectral gap and finally \eqref{eq:comp}. 
 Therefore $\gl_2$ can be chosen equal to $\gl_K$, $C\ge c'_K$,  and therefore
 $M$ is a stable normally hyperbolic manifold in $L^2$ for the reversible Kuramoto
 evolution,  with characteristics 
 $0$, $\gl_K$ and $C= \max(c'_K,1)$. 
 
 %SHOULD WE put back a third (negative) constant to better control the negative continuation?
 \medskip
 
 For the sequel we  observe also that  $u \mapsto P^o_u$, a map from $M$
 to the bounded linear operators on $L^2_0$, is $C^\infty$ as it can be easily
 verified by using for $v \in L^2_0$ the formula
 \begin{equation}
 \label{eq:proj}
 \left( v, q'_\psi\right)_{-1,q_\psi} \, =\, \int_ \bbS \cV
 - \frac{\int_\bbS \cV/q_\psi}{\int 1/q_0}\, ,
 \end{equation}
 where, like before,  $\cV(\theta):= \int_0^\theta v$, so that $\cV:\bbS \to \R$ is (H\"older) continuous
 and $\psi \mapsto  ( v, q'_\psi)_{-1,q_\psi}$ is $C^\infty$.

\subsection{The full evolution equation}
\label{sec:full}

The type of  limit evolution equations we are interested in can be cast into the form
\begin{equation}
\label{eq:main}
\partial_t p^\gd_t (\theta) \, =\, \frac 12 \partial_\theta^2 p^\gd_t(\theta) - \partial_\theta \left[ p^\gd_t(\theta) (J * p_t^\gd)(\theta)
\right] + \gd G[p^\gd_t](\theta)\, , 
\end{equation}
where $\gd \ge 0$ and for $ G$ we assume
\begin{enumerate}
\item $p\mapsto  G[p]$ is a function from $L^2_1$ to $H_{-1}$;
\item there exists $\eta>0$ such that $G$ is $C^1(L^2_1,H_{-1})$ for every $p$ at $L^2$ distance at most $\eta$ of $M$ and 
the derivative $DG$ is uniformly bounded (in the $\eta$-neighborhood of $M$ that we consider).
\end{enumerate}
Note that $p \mapsto (p J*p)'$ is also in $C^1(L^2_1,H_{-1})$, in fact even in $C^\infty$, so that 
the evolution equation can be cast in the abstract form $\partial p_t^\gd =
A p_t^\gd + F[p_t^\gd] + \gd G[ p_t^\gd]$. A complete theory of this type of
equations can be found in  \cite[Ch.~4]{cf:SellYou}, in particular 
 for $p_0^\gd\in L^2_1$ such that $d_{L^2}(p_0^\gd, M) < \eta$
 there exists of a unique mild solution in $C^0([0,T), L^2_0)$, for some $T>0$.  
 
\medskip

Examples  include: 
\begin{enumerate}
\item the AR case, that is 
\eqref{eq:aim}, with $ G[p](\theta)=  \partial_\theta [ p (\theta) U(\theta)]$ and 
 $\Vert U \Vert_\infty < \infty$; 
\item the case of 
\begin{equation}
 G[p](\theta)\, =\,  \partial _\theta [p (\theta) \tilde J * p (\theta)]\, 
\end{equation} 
with $\tilde J\in L^\infty$;
\item the case of
\begin{equation}
G[p](\theta)\, =\,  \partial_\theta [ p(\theta) \int_\bbS h(\theta, \theta') p(\theta') ]\, ,
\end{equation}
with $h\in L^\infty$, as well 
as generalizations like $\partial_\theta [ p(\theta) \int_\bbS h(\theta, \theta', \theta'') p(\theta') p(\theta'') ]$ and so on.
\end{enumerate}

\medskip

In all these examples actually one can prove global well-posedness  for arbitrary
initial condition in $L^2_1$. But the key point of our analysis is that if the initial condition is
sufficiently close to $M$, then for $\gd$ smaller than a suitable constant,  the solution
will stay in a neighborhood of $M$ for all times. More precisely,
our approach is  based on the following result, that is essentially contained 
in \cite[Main Theorem, p.~495]{cf:SellYou}. We say {\sl essentially} because 
the result we need is more explicit for what concerns the  various {\sl small constants}
that are involved:  in Section~\ref{sec:SY} we detail this issue.

\medskip

\begin{theorem}
\label{th:M}
There exists $\gd_0>0$ 
such that if $\gd \in [0,  \gd_0]$ there exists a stable  normally hyperbolic  
manifold  $M_\gd$ in $L^2_1$ for the perturbed equation \eqref{eq:main}. Moreover 
%$M_\gd$ is a $C^1$-manifold in $L^2_0$ and 
we can write 
\begin{equation}
\label{eq:M}
M_\gd \, =\, \left\{q_\psi + \phi_\gd \left(q_\psi\right):\, \psi  \in \bbS \right\}\, ,
\end{equation}
for a suitable  function $\phi_\gd\in C^1(M,L^2_0)$ with the properties that 
\begin{itemize}
\item $\phi_\gd (q) \in \cR (L_q)$;
\item 
there exists 
$C>0$ such that 
$\sup_{\psi }(\Vert \phi _\gd \left(q_\psi\right)\Vert_2+
\Vert \partial_\psi\phi_\gd(q_\psi)\Vert _2) \le C \gd$.
\end{itemize}
\end{theorem}

\medskip

We are now interested in the dynamics on $M_\gd$, which is a curve and, given the mapping $\phi_\gd$,
the position on the manifold is identified by the {\sl phase} $\psi_t^\gd$. A more detailed description 
demands information on $n^\gd_t := \phi_\gd( q_{\psi_t^\gd})$: of course $\psi_t^0=\psi_0^0$ and $n^0_t\equiv 0$
for every $t$. %We assume, without loss of generality, that $\psi^\gd_0=0$. 
%In order to simplify our approach and deal with smooth evolutions we assume that
%\begin{equation}
%\label{eq:smooth}
%\text{condition(s) to be decided}
%\end{equation}
%so that $(t, \theta) \mapsto p^\gd_t (\theta)$ is $C^\infty$, as well as $\psi_t^\gd$ and $y^\gd_t(\theta)$ (...).

We have the following:

\medskip

\begin{theorem}
\label{th:dynM}
For $\gd \in [0, \gd_0]$ we have that $t \mapsto \psi^\gd_t$ is $C^1$ and 
\begin{equation}
\label{eq:dynM1}
\dot{\psi_t^\gd} +\gd \frac{\left( G[q_{\psi_t^\gd}], q^\prime_{\psi_t^\gd}\right)_{-1, 1/q_{\psi_t^\gd}}}{
\left( q', q'\right)_{-1, 1/q}} \, =\, O(\gd^2)\, ,
\end{equation}
with $ O(\gd^2)$ uniform in $t$. Moreover %if we define $n_\psi^\gd=\phi_\gd(q_\psi)$ for all $\psi\in\bbS$ and 
if we call $n_\psi$ the unique solution of 
\begin{equation}
\label{eq:predynM2}
L_{q_\psi} n_\psi \, =\, G[q_\psi]-\frac{\left( G[q_{\psi}], q^\prime_{\psi}\right)_{-1, 1/q_{\psi}}}{
\left( q', q'\right)_{-1, 1/q}}q^\prime_\psi \ \ \ \text{ and } \ \ \
\left(  n_\psi, q'_\psi \right)_{-1,1/ q_\psi} =\, 0\, ,
\end{equation}
we have
\begin{equation}
\label{eq:dynM2}
\sup_\psi \left\Vert 
 \phi_\gd(q_\psi)\, -\, \gd n_\psi \right\Vert_{H_1} \, =\, O(\gd^2)\, .
\end{equation}
\end{theorem}

\medskip

A sharper control on the dynamics on $M_\gd$ can be obtained, under a slightly
stronger assumption on the perturbation $G$: it all boils down to 
go beyond \eqref{eq:dynM1} and for this note that the left-hand side can be written
as $R^\gd (\psi^\gd_t)$ where
\begin{equation}
\label{eq:Reps}
R_\gd (\psi)\, :=\, \frac{
\left(
\left[ \phi_\gd (q_{\psi}) J*\phi_\gd (q_{\psi})  
\right]^\prime
+ \gd \left( 
G\left[q_\psi +  \phi_\gd (q_{\psi})\right] - G\left[q_\psi \right]\right)\, ,\, q'_\psi  
\right)_{-1,1/q_\psi}
}
{(q',q')_{-1,1/q}}\, .
\end{equation}
It is clear that $R_\psi$ is $C^1$, since $\phi^\gd$ is $C^1$. 
\medskip

\begin{theorem}
\label{th:dynMconj}
Under the same assumptions of the previous theorem and 
assuming in addition that $DG$ (recall that $G\in C^1( L^2_1; H_{-1})$)  is uniformly continuous in a $L^2$-neighborhood of 
$M_0$, 
 we have  that 
%\begin{equation}
%\label{eq:dynMconj}
%R^\gd (\psi_t^\gd)\, :=\, 
%\dot{\psi_t^\gd} + \gd \frac{\left( G[q_{\psi_t^\gd}], q^\prime_{\psi_t^\gd}\right)_{-1, 1/q_{\psi_t^\gd}}}{
%\left( q', q'\right)_{-1, 1/q}}\, ,
%\end{equation}
there exists $\gd\mapsto \ell(\gd)$, with
$\ell(\gd)=o(1)$ as $\gd \searrow 0$, such  that %both $\vert R^\gd(\psi)\vert$ and 
\begin{equation}
\sup_{\psi\in \bbS}\vert R^\prime_\gd(\psi)\vert \, \le\, \gd\,  \ell(\gd)\, .
\end{equation}
\end{theorem}

\medskip

\section{Dynamics on $M_\gd$: analysis of the active rotators case}
\label{sec:ARs}
Let us use the results of the previous section to tackle the questions we have raised in the introduction
for the active rotators case
and that, ultimately, boil down to: what is the relation between the 
Isolated  Deterministic one dimensional System $\dot \psi= -V' (\psi)$ (IDS)
and the behavior of the associated $N$ dimensional diffusion, for $N$ large?
So we focus on  \eqref{eq:main} with $G[p]=(pV')'$
% -- we call it ``large scale dynamics" --  
and regularity assumptions on $V'$ are going to appear
along the way. Theorem~\ref{th:M} tells us that if $\Vert V'\Vert_\infty< \infty$, 
at least when $\gd$ is small enough, the $N \to  \infty$ limit system -- ruled by \eqref{eq:main} --
is described by a dynamics on a one dimensional smooth and compact manifold $M_\gd$ equivalent to a circle
and, via Theorem \ref{th:dynM} and Theorem \ref{th:dynMconj}, we have a sharp control on this dynamics.

In order to be precise on this issue let us speed up time by $1/\gd$ 
  in (\ref{eq:dynM1}). If we keep just the leading terms
 we are dealing with the dynamics
\begin{equation}
\label{eq:1st}
\dot \psi \, =\,  - f(\psi)\, , 
\end{equation}
where  $f$ is 
\begin{equation}
\label{eq:approx}
f (\psi)\, : =\,  \frac{\left( G[q_{\psi}], q^\prime_{\psi}\right)_{-1, 1/q_{\psi}}}{
\left( q', q'\right)_{-1, 1/q}}\, .
\end{equation}
We say that $f \in C^1(\mathbb{S}, \mathbb{R})$ -- not necessarily the $f$ in \eqref{eq:approx} -- is generic, or hyperbolic, 
if it has a finite number of zeroes on $\mathbb{S}$ and 
all of them are simple, i.e. for all $\psi $ for which $f(\psi)=0$, we have   $f^\prime(\psi)\neq 0$.
Notice that the set of generic functions is open in $C^1(\mathbb{S}, \mathbb{R})$ and dense:
if the $C^1$ distance of $f$ and $g$  is less than (a constant times) $\epsilon$,   we say  that
the dynamics generated by $f$ and $g$ are $\epsilon$-close. Note that if $\epsilon$ is sufficiently small
then the two dynamics are topologically equivalent. 
 By this we mean 
that there exists a homeomorphism $h:\bbS \to \bbS$ such that $\{ h(\psi(\psi_0, t)):\, t \in \bbR\}$,
where  $\psi(\psi_0, \cdot)$ solves $\dot{\psi}= -f(\psi)$ and $\psi(\psi_0, 0)=\psi_0$ , coincides
with $\{(\phi(h(\psi_0), t):\, t \in \bbR\}$, where $\phi(\phi_0, \cdot)$ solves $\dot{\phi}= -g(\phi)$
and $\phi(\phi_0, 0)=\phi_0$. Moreover we require that $h(\cdot)$ preserves the time orientation,
that is for $a>0$ sufficiently small and $t$, $\vert s\vert \in (0, a]$
we have that $\psi(\psi_0, t)\neq \psi_0 $ and $h(\psi(\psi_0, t))= \phi( h(\psi_0),s)$ imply $s>0$.

 Theorem \ref{th:dynM} and Theorem \ref{th:dynMconj} guarantee therefore that for $\gd$ sufficiently small
 the phase dynamics on the $M_\gd$ manifold speeded up by $\gd^{-1}$
\begin{equation}
\label{eq:phaseMdyn}
 \frac{\dd }{\dd t}{\psi}^\gd_{t/\gd}\, =\, - f(\psi^\gd_{t/\gd})+ \frac 1\gd R_\gd (\psi^\gd_{t/\gd})\, ,
\end{equation}
is $\gd$-close to the dynamics generated by $f(\cdot)$.

The layout of the remainder of this section is, first,  to show 
that even if we fix  $K>1$, by playing on the choice of $V'(\cdot)$, one can generate {\sl arbitrary generic phase dynamics on $M_\gd$}.
In this part we will make also more explicit the link between $V'$ and $f$.
Afterwards, we will work out in detail a few particular cases and expose some {\sl a priori} surprising behaviors, notably that IDS with periodic behavior (active state)
may lead to a $N \to \infty$ dynamics that settles down to a fixed point (quiescent state) or that IDS without periodic 
behavior may give origin to periodic $N \to \infty$ behaviors. 

\subsection{Noise and interaction induce arbitrary generic dynamics}
\label{sec:gendyn}

It is practical and sufficient to work with $V'(\cdot)$ that is a trigonometric polynomial, that is
\begin{equation}
\label{eq:poly_potential}
V'(\theta)\, =\, a_0+\sum_{j=1}^n \left( a_j\cos(j\theta)+b_j\sin(j\theta)\right)\, .
\end{equation}

\medskip

\begin{theorem}
\label{th:equdyn}
For any generic dynamics on the circle $\dot \psi_t=-f(\psi_t)$ with $f \in C^{1}(\mathbb{S}; \bbR)$
and for any value of $K>1$ there exists a trigonometric polynomial 
 $V'(\cdot)$ (see Remark~\ref{rem:explicit} for an explicit expression) 
such that for $\gd$ small enough, the phase dynamics on $M_\gd$ \eqref{eq:phaseMdyn} 
is $\gd$-close to  
$\dot \psi=-f^\prime(\psi)$. 
\end{theorem}

\medskip
\noindent
\textit{Proof.}
Let $f$ be a generic function in $C^1$. 
By the Stone-Weierstrass Theorem, for every $\gep>0$ there exists a trigonometric
polynomial $P(\cdot)$ such that $\Vert f' -P\Vert_\infty \le  \gep$.
If $c_0$ is such that $\int_0^{2\pi } (P-c_0)=0$ then, since $\int_0^{2\pi} f'=0$,
$\vert c_0 \vert \le \gep$. Thus if we define the trigonometric polynomial
 $Q(\psi):= f(0)+\int_0^\psi (P(\theta) -c_0) \dd \theta$ we have
\begin{equation}
\Vert Q -f\Vert _{C_1}\,  =\,
\Vert Q-f \Vert_\infty +\Vert P-c_0-f ' \Vert_\infty\, \le \, (2\pi+1)   \Vert P-c_0-f ' \Vert_\infty
\, \le \, (4\pi +2) \gep\, , 
\end{equation}
 so it suffices to consider functions $f$ which are trigonometric polynomials:
 \begin{equation}
\label{eq:poly_f}
f(\theta)\, =\, A_0+\sum_{k=1}^n \left( A_k\cos(k\theta)+B_k\sin(k\theta)\right)\, .
\end{equation}
Now we observe that 
if $V'(\cdot)$ is of the form \eqref{eq:poly_potential}
then a straightforward calculation gives
\begin{equation}
\label{eq:scalar product}
 \frac{\left( G[q_{\psi}], q^\prime_{\psi}\right)_{-1, 1/q_{\psi}}}{
\left( q', q'\right)_{-1, 1/q}} \, =\, a_0+ \frac{I_0}{I_0^2-1}\sum_{k=1}^n 
\left(
I_ka_k\cos(k\psi)+I_kb_k\sin(k\psi) \right) \, ,
\end{equation}
where 
\begin{equation}
\label{eq:agt5}
 I_k\,=\,I_k(2Kr(K)) :=\, \frac{1}{2\pi}\int_0^{2\pi}\cos(k\theta)e^{2Kr(K)\cos(\theta)}d\theta\, .
\end{equation}
Therefore by making the choice $a_0:=A_0$ and for $k =1,2, \ldots, n$
\begin{equation}
\label{eq:agt6}
a_k\, :=\, \frac{I_0^2-1}{I_0 I_k}A_k\ \ \  \text{ and } \  \ \ b_k\, :=\, 
\frac{I_0^2-1}{I_0 I_k} B_k\, ,  
\end{equation}
we obtain the function $V'(\cdot)$ we were after. 
\qed

\medskip

\begin{rem}
\label{rem:explicit}
\rm 
The link between $f$ and $V'$ can be made more explicit.
 In fact from
\eqref{eq:agt5} and \eqref{eq:agt6} 
and the fact that the Fourier series of $q_0$ is 
\begin{equation}
\label{eq:q0-Fourier}
q_0(\psi) \, =\,  \frac{1}{2 \pi I_0(2Kr)} e^{2Kr \cos(\psi)} \, =\, \frac{1}{2\pi} + \frac{1}{\pi} \sum_{j=1}^{+ \infty} \frac{I_j(2Kr)}{I_0(2Kr)} \cos(j \psi)\, ,
\end{equation}
one directly extracts 
that
\begin{equation} 
\label{eq:dhs}
f  \, = a_0 + \frac{ I_{0}(2Kr(K))^2}{I_0(2Kr(K))^2 -1} \left ( q_0 \ast V^\prime - a_0 \right )\, =\, 
a_0 + D(K) q_0 \ast (V'-a_0)
                        \, ,
\end{equation}
where we have set 
\begin{equation}
\label{eq:J}
 D(K):= \frac{I_0^2(2Kr(K))}{(I_0^2(2Kr(K))-1)}\, ,
\end{equation}
and \eqref{eq:dhs}
 can be applied also in the case in which $f$ is not
a trigonometric polynomial. It tells us that, for $\gd$ small, the {\sl effective force} that drives
the $N \to \infty$ system is, in a sense, obtained by smearing $V'$ via the probability kernel
$q_0$. To be precise,  $V'-a_0$   is smeared and multiplied by $D(K)$, while 
the $0^{\textrm{th}}$ order Fourier coefficient is left unchanged.  This is telling us that the effect
of noise and interaction, to leading order, boil down to the size of  
$D(K)$ and to the smearing effect of the probability kernel $q_0(\cdot)$ (that depends on $K$ too!). 
\end{rem}
%%%%%%%%%%%%%%%%%%%%%%%%%%%%%%%%%%%%%%%%%%%%%%%%%%%%%%%%%%%%%%%%%%%%%%%%%%%%%%%%%%%%%%%%%%%%%%%%%%%%%%%%%%%%%%%%%%%%%%%%%%%%%%%%%%%%%%%%%%%%%%%%%%%%%%%%%%%%%%%%

%A striking consequence of this is that there is a range of $K$, independent of potential $V$, such that  for any given not decreasing (and not constant) potential $V$, for $\delta$ small enough,  the collective dynamics described by equation \eqref{eq:main} is periodic on the one dimensional invariant curve $M_\delta$. This holds when for any $k \geq 1$, $\frac{A_k}{a_k} < 1$ and $\frac{B_k}{b_k} < 1$, ie  as soon as $\frac{I_0(2Kr(K)) I_k(2Kr(K))}{I_0^2(2Kr(K)) -1} < 1$ for any $k \geq 1$, and even if $V$ is constant on some intervall of $\mathbb{S}$. 

%%%%%%%%%%%%%%%%%%%%%%%%%%%%%%%%%%%%%%%%%%%%%%%%%%%%%%%%%%%%%%%%%%%%%%%%%%%%%%%%%%%%%%%%%%%%%%%%%%%%%%%%%%%%%%%%%%%%%%%%%%%%%%%%%%%%%%%%%%%%%%%%%%%%%%%%%%%%%%%%

\medskip

While \eqref{eq:dhs} is quite explicit, it is not always straightforward to read off it
the qualitative properties of $f$. We start by analyzing the case of $K$ very large and the case 
of $K$ close to one, before moving to treating in detail some particular cases.

\subsubsection{The $K \to \infty$ limit}
It is  straightforward to see that the probability density $q_0(\cdot)$ converges
to the Dirac delta measure at the origin. Moreover $\lim_{K \to \infty} D(K) =1$,
since $\lim_{K\to \infty}r(K)=1$ and $\lim_{x \to \infty}I_0(x)=\infty$. Therefore
$f$ and $V'$ get closer and closer as $K$ becomes large. More precisely one 
has that for every $s\in \bbN$ and every trigonometric polynomial  $V'(\cdot)$ there
is $C$ such that
\begin{equation}
\label{eq:CK}
\Vert f -V' \Vert_{C^s}\, \le \, \frac C{\sqrt{K}} \, .
\end{equation}
The proof can be obtained for example by using \eqref{eq:agt6} that, with \eqref{eq:J}, tells us
that $A_j/a_j$, as well as $B_j/b_j$, that is the ratio of the (non-vanishing) sine and cosine  Fourier coefficients 
of $f$ and $V'$, is 
\begin{equation}
\label{eq:ratioJ}
D(K) \frac{I_j(2Kr(K))} {I_0(2Kr(K))}\, ,
\end{equation}
%(even) Fourier coefficients of the (even) function $q_0$ are given by the modified Bessel functions: 
%\begin{equation}
%\label{eq:qcoeff}\hat{q_0}(j) \, =\,  \frac{I_j(2Kr)}{I_0(2Kr)}\ , \end{equation}
%which is just a restatement of \eqref{eq:q0-Fourier}
so that
by using  $(I_j(x)/I_0(x))-1 \stackrel{x \to  \infty}\sim \frac{j^2}{2x}$  ($j=1,2, \ldots$)
and $\lim_{K \to \infty}D(K)=1$ we readily obtain
 that the $j^{\textrm{th}}$-Fourier coefficients of $f(\cdot)$
%of $q_0 \ast(V'-a_0)$ (recall \eqref{eq:dhs}) 
are, to leading order, $ j^2 a_j /(4K)$ and $ j^2 b_j /(4K)$.
Since
we are just dealing with trigonometric polynomials and the estimate of
the $L^2$ norm of arbitrary derivatives of $f-V'$, via Parseval formula, is straightforward, we get to \eqref{eq:CK}.
%%%%%%%%%%%%%%%%%%%%%%%%%%%%%%%%%%%%%%%%%%%%%%%%%%%%%%%%%%%%%%%%%%%%%%%%%%%%%%%%%%%%%%%%%%%%%%%%%%%%%%%%%%%%%%%%%%%%%%%%%%%%%%%%%%%%%%%%%%%%%%%%%%%%%%%%%%%%%%%%%%%%%%%%
This means in particular that given a potential $V$ such that $V^\prime$ has sign changes (so that the IDS
has stable points), 
for any $K$ large enough, the $N\to \infty$  system  
has stable stationary solutions, for $\delta$ small enough.
We will encounter this phenomenon in the particular cases that we treat below.  
%%%%%%%%%%%%%%%%%%%%%%%%%%%%%%%%%%%%%%%%%%%%%%%%%%%%%%%%%%%%%%%%%%%%%%%%%%%%%%%%%%%%%%%%%%%%%%%%%%%%%%%%%%%%%%%%%%%%%%%%%%%%%%%%%%%%%%%%%%%%%%%%%%%%%%%%%%%%%%%%%%%%%%%%

\subsubsection{The $K \searrow 1$ limit}
This time we use $r(K)\stackrel{K \searrow 1}\sim \sqrt{2(K-1)}$ and we derive, first of all,
that $D(K) \sim (4(K-1))^{-1}$, since $I_0(x)-1 \stackrel{x\searrow 0}\sim x^2/4$.
Once again we analyze the Fourier coefficients of $f$, via \eqref{eq:ratioJ},
and  we use for $j=1,2, \ldots$
\begin{equation}
\frac{I_j(x)}{I_0(x)}  \stackrel{x \searrow 0}{\sim} I_j(x) \, \sim\,
\frac{x^j}{2^j j!}\, ,
\end{equation}
so that for $j=1,2, \ldots$
\begin{equation}
\frac{A_j}{a_j}\, =\, \frac{B_j}{b_j}  \stackrel{K \searrow 1}{\sim}
\frac{(K-1)^{-1+(j/2)}}{2^{2-(j/2)}j!}\, .
\end{equation}
Notably, the first Fourier coefficients of $f$ are enhanced with respect to the corresponding 
 coefficients of $V'$ by a factor that diverges like $(K-1)^{-1/2}$. The second Fourier coefficients
of $f$ are (asymptotically) just proportional to the ones of $V'$, while higher coefficients in the
$K \searrow 1$ limit
are depressed passing from $IDS$ to $N\to \infty$ behavior (recall that the $0^{\textrm{th}}$-order coefficient
is unchanged). A quantitative estimate in the
spirit of \eqref{eq:CK} is easily established from these estimates.

What we retain from this $K \searrow 1$ analysis is that if the first Fourier coefficients are present, that is
$\vert a_1\vert+\vert b_1 \vert >0$, then for $K$ sufficiently close to one $f(\psi)=0$ has
two solutions and the dynamics will eventually settle to a fixed point (quiescent state). If instead 
$\vert a_1\vert+\vert b_1 \vert =0$, then it depends on the relative size of $a_0$ and $a_2$ or $b_2$
whether the system is in an activated or quiescent regime.
But if also $a_2=b_2=0$ (and $a_0\neq 0$) then for $K$ sufficiently close to one we have that 
$f(\psi)$ is close to $a_0$ and therefore $f(\psi)\neq 0$ for all $\psi$, so that the dynamics is periodic. 
Again, we will discuss in more detail these issues below, in specific examples.

%%%%%%%%%%%%%%%%%%%%%%%%%%%%

\medskip
\begin{rem}
\rm
The analysis for $K$ large and close to one is helpful to get an idea on the relation
between $f$ and $V'$, but the reader should keep in mind that the
$\gd$-closeness of the dynamics holds for fixed $K$, that is for $\gd < \gd_0(K)$. Quantitative estimates on
how $\gd_0(K)$ behaves for extreme values of $K$ is an interesting issue that we do not 
approach here.
\end{rem}

\subsection{Active rotators with $V(\theta)=\theta-a\cos(\theta)$}
Without loss of generality we assume $a\ge 0$.
Let us start the analysis by making a remark on the 
$a=0$ case: the potential becomes just a straight line, and \eqref{eq:main} reads
\begin{equation}
 \partial_t p^\gd_t (\theta) \, =\, \frac 12 \partial_\theta^2 p^\gd_t(\theta) - \partial_\theta \left[ p^\gd_t(\theta) (J * p_t^\gd)(\theta)-\gd p^\gd_t(\theta)
\right] \, .
\end{equation}
In this case $p_t^\gd(\theta-\gd t)$ solves \eqref{eq:Kuram}, thus  $M_\gd=M$ and the dynamics
on $M_\gd$ is a rotation for all $\gd$.

\begin{figure}[hlt]
\begin{center}
\leavevmode
\epsfxsize =14 cm
\psfragscanon
\psfrag{1/mem}{$a_c(K)$}
\psfrag{v}{$K$}
\psfrag{formula}{$a_c(K)\, =\, \frac{I_0^2(2Kr) -1}{I_0(2Kr)I_1(2Kr)}$
\ and \  $r\, =\,  \frac{I_1(2Kr)}{I_0(2Kr) }$}
\psfrag{th}{$\theta$}
\psfrag{pi}{$\pi$}
\psfrag{-pi}{$-\pi$}
\psfrag{0}{$0$}
\epsfbox{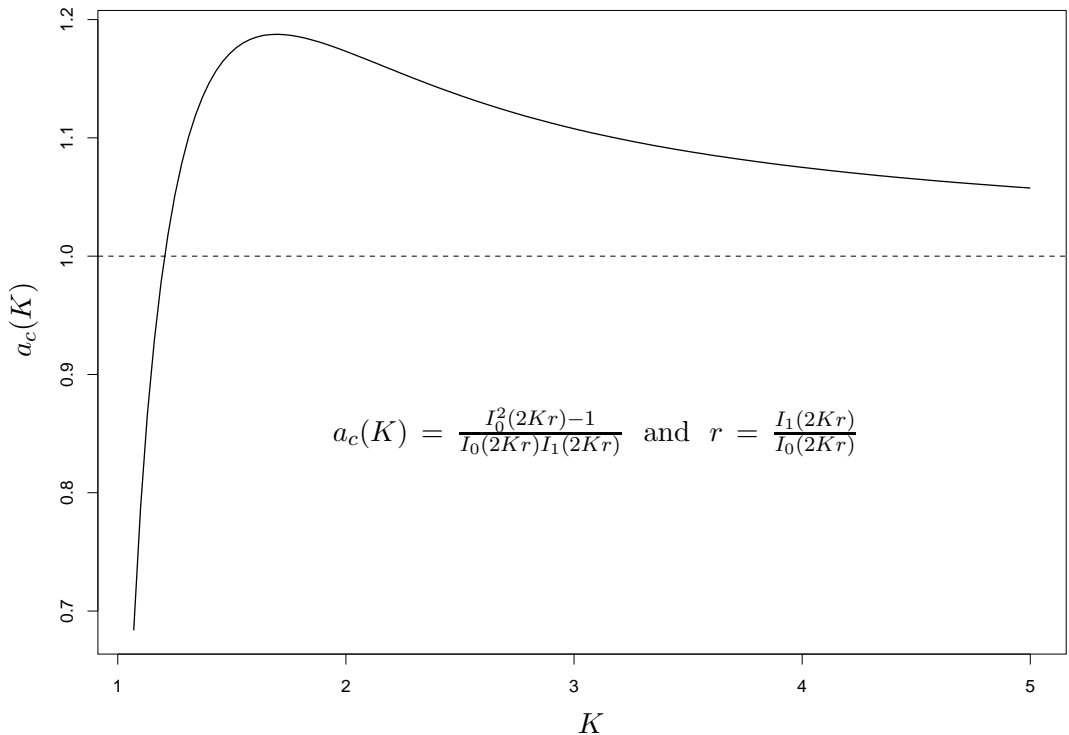}
\end{center}
\caption{The graph of $a_c(\cdot)$. For $K \to \infty$ we have
$a_c(K)\, =\, 1+ 1/(8K)+O(K^{-2})$,
while for $K \searrow 1$ we have
$a_c(K)\, \sim\,  \sqrt{32 (K-1)} $.
}
\label{fig:W}
\end{figure}

\begin{figure}[hlt]
\begin{center}
\leavevmode
\epsfxsize =12 cm
\psfragscanon
\psfrag{W}{$a_c(K)$}
\psfrag{a}{$a$}
\psfrag{0}{$0$}
\epsfbox{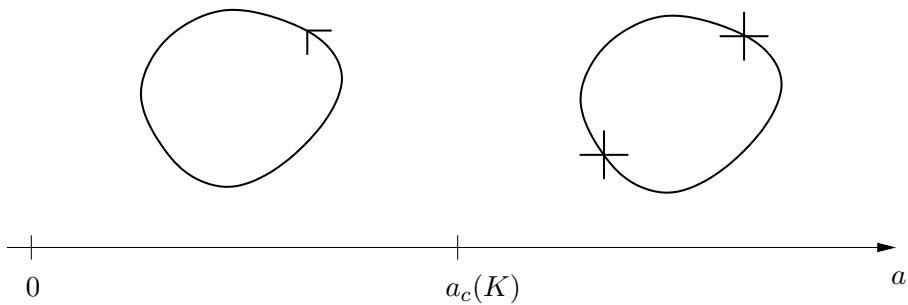}
\end{center}
\caption{A sketch of the phase behavior for $V(\theta)=\theta-a\cos(\theta)$: for $a>K$ there 
are two fixed points, one attractive and one repulsive, while for $a<a_c(K)$ the force is bounded away from zero
and the motion is periodic.}
\label{fig:W2}
\end{figure}

If $a>0$ we exploit the analysis we have developed for 
Theorem~\ref{th:equdyn} that tells us that the $N\to \infty$ phase dynamics 
is lead by the effective force
\begin{equation}
\label{eq:dynMpc}
 f(\psi)\, =\, - \left(1+\frac{a}{a_c(K)}\sin(\psi)\right)\, ,  \ \ \text{ with } \ \ a_c(K)\, :=\, \frac{I_0^2-1}{I_0I_1}\, . 
\end{equation}
Therefore 
if $a<a_c(K)$, then the dynamic on $M_\gd$ is periodic for $\gd$ small enough ( depending on $K$ )
and  if $a>a_c(K)$,  there are two fixed points. From this observation and the graph of $a_c(\cdot)$
(see Figure~\ref{fig:W}) we draw the following conclusions (see also Figure~\ref{fig:W2}):
\begin{itemize}
\item Set $\hat a_c:=\max_K a_c(K)( >1)$. If $a>\hat a_c$ then for every $K$ we have that $f(\theta)=0$ has two solutions,
so that the phase dynamics has two stationary hyperbolic point: one is stable and the other is unstable.
In this case the dynamics of the IDS resembles to the phase dynamics of the $N \to\infty$ system.
\item If $a\in (1,\hat a_c)$ then $a_c(K)=a$ has two solutions $K_-(a)< K_+(a)$ and for 
$K\in (K_-(a), K_+(a))$ we have $a< a_c(K)$, that is $f(\theta)<0 $ for every $\theta$, and  
the motion is periodic: in this case the dynamics of the IDS, that has two fixed points, differs from the $N \to \infty$ phase dynamics.  For $K >K_+(a)$ and for $K< K_-(a)$ instead the phase dynamics is driven
to a (unique) stable fixed point (unless it starts from the unstable fixed point). 
\item If $a\le 1$ instead $a_c(K)=a$ has only one solution $K(a)$ and the periodic behavior sets up for
$K>K(a)$, otherwise ($K<K(a)$) the system eventually settles on a fixed point: this second case is another instance in which
the dynamics of the IDS and the $N \to \infty$ system differ. 
\end{itemize}

When the phase dynamics is periodic we can explicitly integrate the evolution equation
\eqref{eq:1st} and compute the first order approximation 
the period $T_\gd (a,K)$ of the dynamics on $M_\gd$:
\begin{equation}
\label{eq:TaK}
T_\gd (a,K)\, =\, \frac {\tau(a,K)}{\gd}
+O(1)\, , \ \ \text{ where } \ \  \tau(a,K)\, :=\, 
 \frac{2\pi}{\sqrt{1-(a /a_c(K))^2}} \, .
\end{equation}
Actually, it is possible to replace in this formula $O(1)$ with $O(\gd)$: in fact  it is possible to show by induction that the phase speed on $M_\gd$ admits an expansion in (integer) powers of $\gd$
 to any order (but with coefficients less explicit than the first order one), and it is easy to see that $\dot{\psi}_\gd$ is an odd function of $\gd$. 
We have tested numerically this approximation and we report the result 
in Table~\ref{tab:1}.

\begin{SCtable}[45][tbp]
%\begin{wide}
  \begin{tabular}{ | l | r | r | }
    \hline \phantom{m}
    $\gd$& $T_\gd(1.1,2)$ 
     & $\tau(1.1,2)/ \gd   $  \\ \hline
     0.005 & 3615.59 &	 3615.62 \\ \hline
 0.010 & 1807.79 & 1807.85 \\ \hline
 0.020 &  903.89 &   904.01 \\ \hline
 0.040  & 451.94 & 	452.19 \\ \hline
 0.080  & 225.97 & 	226.45 \\ \hline
0.160  & 112.98 & 	113.96 \\ \hline
0.320   & 56.49 & 	 58.51 \\ \hline
0.640   & 28.24 & 	 33.02 \\ 
    \hline
   \end{tabular}
  \caption{
  We have simulated  \eqref{eq:aim} with $V(\theta)=\theta-a \cos(\theta)$
  for $a=1.1$ and
  $K=2$. 
  In this case our estimates ensure the existence of periodic solutions 
  for $\gd$ sufficiently small and the period given in \eqref{eq:TaK} (in fact, 
   $\tau:=\tau(1.1,2)= 18.0779\ldots$). The 
   simulation,  that has been performed via Fourier decomposition (50 modes kept), 
  gives $c=0.333\ldots $, for the constant $c$   such that $(\gd T_\gd(1.1,2)/\tau (1.1,2) )-1\sim c \gd^2$.
  %, for the chosen values of $a$ and $K$.
   \label{tab:1}}
 %   \end{wide}
\end{SCtable}

\subsection{Active rotators with $V(\theta)=\theta -a\cos(j\theta)/j$, $j=2,3, \ldots$}
\begin{figure}[hlt]
\begin{center}
\leavevmode
\epsfxsize =14 cm
\psfragscanon
\psfrag{x}{$K$}
\psfrag{y}{$a_{c,2}(K)$}
\psfrag{y3}{$a_{c,3}(K)$}
\epsfbox{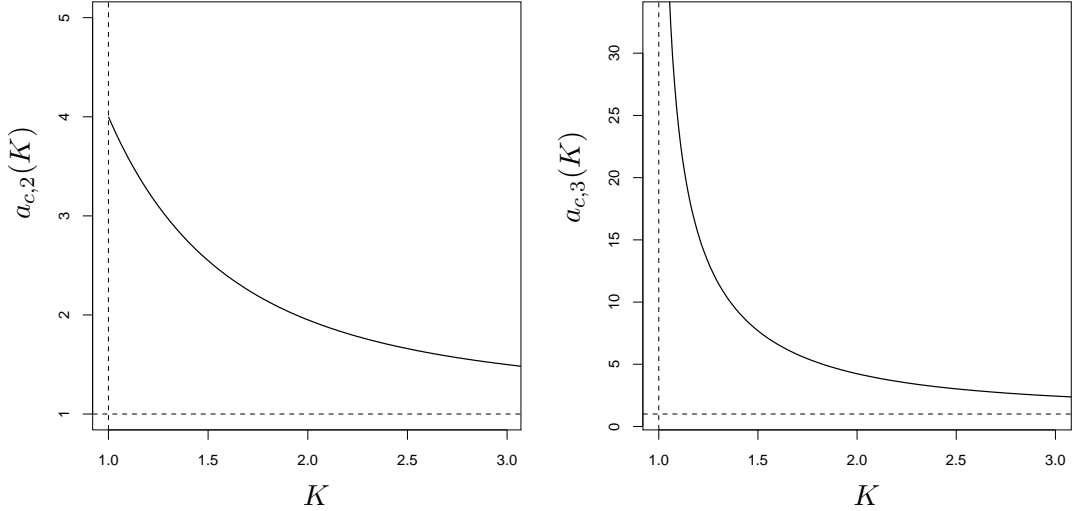}
\end{center}
\caption{For $V(\theta)=\theta -a\cos(j\theta)/j$, $j\ge 2$, the $N \to \infty$ dynamics is always 
periodic for $a\le 1$ (and $\gd $ sufficiently small), unlike the $j=1$ case (recall that
$a<a_{c,j}(K)$ corresponds to periodic motion, while $a>a_{c,j}(K)$ corresponds to two fixed points: see the text).  Moreover, for
$j=2$ and $a>4$ the dynamics has just two fixed points, but for $j\ge 3$ for arbitrarily
large values of $a$ one can observe periodic motion if $K$ is sufficiently close to $1$
(and, of course, $\gd$ sufficiently small).}
\label{fig:j}
\end{figure}

In this case the $N \to \infty$ phase dynamics is lead by 
\begin{equation}
 f(\psi)\, =\, -\left(1+ a\frac{I_0I_j}{I_0^2-1} \sin(j\psi)\right)\, ,
\end{equation}
and the behavior differs substantially from the $j=1$ case (and the $j=2$ case
is different from the $j\ge 3$ case).
In this case the crucial function is 
\begin{equation}
a_{c,j}(K)\, :=\, \frac{I_0^2-1}{ I_0 I_j}\, .
\end{equation} 
Note that $a_{c,1}=a_c$. The criterion to have periodic behavior is, like for the $j=1$ case, 
$a< a_{c,j}(K)$, while $a> a_{c,j}(K)$ leads to two fixed points.
Figure~\ref{fig:j} and its caption describes the (relatively surprising) phenomenology of the
$j=2$ and $j=3$ cases (the case $j>3$ is qualitatively the same as the case $j=3$).

\begin{rem}\rm
Theorem~\ref{th:equdyn} already tells us that one can produce arbitrary dynamics, so 
a very large variety of phenomena is observed. 
Here is a case that can be of some interest since it shows that playing on  only one parameter 
one can produce three different dynamics (and the reader will directly infer how to induce
arbitrarily many):
if  $V(\theta)=\theta -a(\cos(\theta)+\cos(2\theta))$
 the $N \to \infty$ phase dynamics is lead by 
\begin{equation}
 f(\psi)\, =\,1+a\frac{I_0}{I_0^2-1}\left(I_1\sin(\psi)+2 I_2\sin(2\psi)\right)\, ,
\end{equation}
and in this case there can be two transitions as $a$ varies. For example
 for $K=2$ we have periodic behavior for $a<0.600\ldots$, 
two fixed points if $a \in (0.600\ldots , 2.107\ldots)$ and four fixed points (of course two stable
and two unstable ones) if $a>2.107\ldots$.
\end{rem}

\section{Perturbation arguments}
\label{sec:epsilon}

In this section 
we assume   that $\gd \in (0,\gd _0]$ (cf. Theorem~\ref{th:M})
and that we are on the invariant manifold $M_\gd$ of \eqref{eq:main}, that is $p^\gd_t \in M_\gd$ for every $t$. 
The result
\cite[Main Theorem, p.~495]{cf:SellYou} actually contains also some estimates on the regularity
of the semigroup on $M_\gd$ and notably that $t\mapsto p^\gd_t$
belongs to $C^0(\bbR ; \tilde H_1)$ and that it 
 is (strongly) differentiable
as a map from $\bbR$ to $\tilde H_{-1}$. One directly sees that
$\Vert u-v \Vert_{H_1}= \Vert u''-v'' \Vert_{-1}$, so that 
the right-hand side in \eqref{eq:main} is $C^0(\bbR; H_{-1})$ and, in turn, 
$t\mapsto p^\gd_t$ is $C^1(\bbR,\tilde H_{-1})$.

\medskip

Since we are working in a neighborhood of $M$ it is useful to introduce from now a parametrization
of this region that will be particularly useful in the next section, but that we are going to use from now.
The following facts are proven in Lemma \ref{lem:parametrisation}:
for every $u$ in a sufficiently small $H_{-1}$ neighborhood of $M$ there exists a unique $q=v(u)\in M$
such that 
\begin{equation}
\left( u-q, q' \right)_{-1,1/q}\, =\, 0\, .
\end{equation}
Furthemore $v \in C^1 (\tilde H_{-1}, \tilde H_{-1})$ with differential
\begin{equation}
\label{eq:Dv}
Dv(u) \, =\, P^o_{v(u)}\, .
\end{equation}
 Theorem~\ref{th:M} is telling us in particular that
\begin{equation}
v\left( q + \phi_\gd \left(q\right) \right) \,= \, q \, .
\end{equation}
For the arguments that follow it is practical to use the notation introduced right after
Theorem~\ref{th:M} and write 
\begin{equation}
\label{eq:decomp}
p^\gd_t\, =\, q_{\psi^\gd_t}+ n^\gd_{t}\, ,
\end{equation}
where $ q_{\psi^\gd_t}= v(p^\gd_t)$
and $n_t^\gd := \phi_\gd \left( q_{\psi^\gd_t}\right)$.

\medskip

\noindent
{\it Proof of Theorem~\ref{th:dynM}.}
Since the evolution on $M_\gd$ is $C^1(\bbR,\tilde H_{-1})$,
then $t \mapsto q_{\psi^\gd_t}$ is $C^1(\bbR,\tilde H_{-1})$ too.
This implies  that, with $f_1$ and $f_2$
respectively  sine and cosine, $\psi\mapsto \int_\bbS q_{\psi}(\theta) f_i (\theta)=: a_i(t)$
is $C^1$.  Since $f_i(\psi_t)= a_i(t)/\sqrt{a_1^2(t)+a_2^2(t)}$, we see
that $t\mapsto \psi^\gd_t$ is $C^1$. The fact that $p^\gd_t$ and $\psi^\gd_\cdot$ are $C^1$ directly implies 
 that $t\mapsto n_t^\gd$ is $C^1(\bbR; H_{-1})$ (actually, since $\phi_\gd$ is $C^1$ we have even
 $n_\cdot ^\gd \in C^1(\bbR; L^2_0)$).

Notice furthermore that
\begin{equation}
\label{eq:linkpsip}
-\dot{\psi^\gd_t}q^\prime_{\psi^\gd_t} \, =\, P^o_{q_{\psi^\gd_t}}\partial_t p^\gd_t\, .
\end{equation}
This follows by taking the time derivative of both sides of the equality
$q_{\psi^\gd_t}= v(p^\gd_t)$ and by using \eqref{eq:Dv}. 

Using \eqref{eq:main} and the fact that $q_{\psi}$ is a stationary solution of \eqref{eq:Kuram}, we rewrite \eqref{eq:linkpsip} as
\begin{equation}
\label{eq:projdynM}
-\dot{\psi^\gd_t} q^\prime_{\psi^\gd_t}\, = \, P^o_{q_{\psi^\gd_t}}\left(-\partial_\theta \left[ n^\gd_{t} (J * n^\gd_{t})
\right]+\gd G\left[q_{\psi^\gd_t}+n^\gd_{t}\right]\right)\, .
\end{equation}
Recall that
$\left\Vert n_t^\gd\right\Vert_2 \le C \gd$ (cf. Theorem~\ref{th:M}): by 
\begin{equation}
\label{eq:boundJ}
\Vert  \left[ n^\gd_t J* n^\gd_t \right]' \Vert_{-1}\, \le \,
\Vert J \Vert_2 \Vert n^\gd_t\Vert_2^2\, \le \, C^2\Vert J \Vert_2^2 \gd^2 \, ,
\end{equation}
and 
by the hypothesis on $G$  that implies that 
\begin{equation}
\label{eq:boundG}
\left \Vert 
G\left[ q_{\psi_t^\gd} + n^\gd_t \right]- G\left[ q_{\psi_t^\gd} \right]
\right \Vert_{-1}\, \le\, c_G C \gd \, , 
\end{equation}
from \eqref{eq:projdynM}  
we see that
\begin{equation}
\label{eq:fmain2}
\left \Vert \dot{\psi_t^\gd} q'_{\psi_t^\gd} 
+ \gd G\left[ q_{\psi_t^\gd}  \right] \right\Vert_{-1} \le \, c \gd ^2\, , 
\end{equation}
with $c$ independent of $t$ and of $\psi_0^\gd$. 
To obtain \eqref{eq:dynM1} just 
take the $H_{-1, q_{\psi_t^\gd}}$ scalar product of $q'_{\psi_t^\gd}$ 
and  the expression inside the norm in the left-hand side of \eqref{eq:fmain2}.

For \eqref{eq:dynM2} rewrite \eqref{eq:main} as
\begin{equation}
\label{eq:main2}
-\dot{\psi_t^\gd} q'_{\psi_t^\gd} - \partial_t n^\gd_{t}\, =\, 
-L_{q_{\psi_t^\gd}} n_{t}^\gd - \left[ n^\gd_{t} J* n^\gd_{t} \right]' + \gd G\left[ q_{\psi_t^\gd} + n^\gd_{t} \right]\, .
\end{equation}
Note that for the second term on the left hand side we have 
%\begin{equation}
%\Vert n^\gd_{{t_1}}- n^\gd_{{t_2}}
%\Vert_2 \, =\, \left\Vert \phi_\gd \left( q_{\psi^\gd_{t_1 }}\right)-
%\phi_\gd \left( q_{\psi^\gd_{t_2 }}\right)\right\Vert_2 \, \le \, C \gd \vert \psi^\gd _{t_1}-\psi^\gd_{t_2}\vert \, , 
%\end{equation}
%where the last inequality holds for $\vert t_1-t_2\vert$ sufficiently small, so small that we can apply the 
%estimate on the Lipschitz constant of $\phi_\gd (\cdot)$ in Theorem~\ref{th:M}.
\begin{equation}
\label{eq:boundL}
\Vert 
\partial_t n^\gd_{\psi^\gd_{t}}\Vert_{-1} \,  \le \, c_K \Vert \partial_t n^\gd_{\psi^\gd_{t}}\Vert_{2}
\,  \le \, c_K
 C \gd \vert \dot{\psi_t^\gd}\vert \, ,
\end{equation}
where we have use
\begin{equation}
 \partial_t n^\gd_{t}\, =\,  \dot{\psi_t^\gd}  \partial_\psi \phi_\gd \left(q_{\psi}\right)
 \big \vert_{\psi=\psi_t^\gd}\, ,
\end{equation} 
and the bound on the derivative of $\phi_\gd$ given in Theorem~\ref{th:M}.

Now plug \eqref{eq:dynM1} into \eqref{eq:main2} and use 
 \eqref{eq:boundJ}, \eqref{eq:boundG} and  \eqref{eq:boundL}  to obtain 
 \begin{equation}
\label{eq:fordynM2}
\sup_{t,\psi_0^\gd} \left\Vert 
 L_{q_{\psi_t^\gd}} n_{\psi^\gd_t}\, -\, \gd \left(G\left[q_{\psi_t^\gd}\right]-\frac{\left( G\left[q_{\psi_t^\gd}\right], q^\prime_{\psi_t^\gd}\right)_{-1, 1/q_{\psi_t^\gd}}}{
\left( q', q'\right)_{-1, 1/q}}q^\prime_{\psi_t^\gd}\right)\right\Vert_{-1} \, =\, O(\gd^2)\, .
\end{equation}
Since $\psi_0^\gd$ can be chosen arbitrarily on $\bbS$, we can replace $\psi_t^\gd$  with $\psi$
and take the supremum over $\psi$ (and, by \eqref{eq:equiv}, we can freely switch between
$H_{-1}$ and $H_{-1, 1/{q_\psi}}$ norms). Therefore (recall \eqref{eq:predynM2})
\begin{equation}
\sup_\psi 
\left\Vert 
 L_{q_{\psi}} \left(n_\psi^\gd\, - \gd n_\psi \right)\,
 \right\Vert_{-1,1/q_\psi} \, =\, O(\gd^2)\, .
\end{equation}
There result we are after, that is  \eqref{eq:dynM2}, follows from the equivalence of
$H_1$ and $V^2_q$ (recall \eqref{eq:Vq}) norms, which is proven in Appendix~\ref{app:normeq}.
\qed

%\begin{rem}\rm
%One could improve this result  by establishing certain norm equivalences.What is clear is that since $n_\psi^\gd\, - \gd n_\psi$ is orthogonal to $q'_\psi$ we have 
%\begin{equation} \left\Vert  L_{q_{\psi}} \left(n_\psi^\gd\, - \gd n_\psi \right)\,\right\Vert_{-1,1/q_\psi}  \, \ge \,  \gl_K \left\Vert  n_\psi^\gd\, - \gd n_\psi \, \right\Vert_{-1,1/q_\psi} ,\end{equation}
%and therefore, since $(u, (1+L_q) u)_{-1,1/q}\ge c (u,u)_2$ for a suitable positive constant $c$,we have
%\begin{equation}\left\Vert  L_{q_{\psi}} \left(n_\psi^\gd\, - \gd n_\psi \right)\, \right\Vert_{-1,1/q_\psi}  \, \ge \,  \frac{c}{{1+1/\gl_K}} \left\Vert   \sqrt{L} \left(n_\psi^\gd\, - \gd n_\psi \right) \right\Vert_{2} \, . \end{equation}.
%\end{rem}

\medskip

\noindent
{\it Proof of Theorem~\ref{th:dynMconj}}.
It is of course sufficient to estimate the numerator in the right-hand side of \eqref{eq:Reps}.
It is the sum of two terms:
the first one can be rewritten as 
\begin{equation}
\label{eq:Reps1}
T_1(\psi)\, :=\, 
\int_\bbS
 \phi_\gd (q_{\psi}) J*\phi_\gd (q_{\psi})  \left(1-  \frac{2\pi/q_\psi}{\int_\bbS 1/q}
 \right)\, ,
\end{equation}
and, by derivating and using the two $L^2$-estimates on $\phi_\gd(\cdot)$ and $D\phi^\gd(\cdot)$ in Theorem~\ref{th:M},  it is straightforward to see that there exists $c>0$ such that for $\gd \in [0, \gd_0]$
\begin{equation}
\label{eq:T1est}
\sup_{\psi\in \bbS}  \vert T^\prime_1(\psi)\vert \, \le \, c \gd^2 \, .
\end{equation}
Let us turn to the second term, that is
\begin{equation}
 T_2(\psi)\, =\, \int_\bbS \left(1-\frac{2\pi /q_\psi(\theta)}{\int 1/q}\right)\int_0^\theta \left(G[q_\psi(\theta^\prime)+\phi^\gd(q_\psi(\theta^\prime))]-G[q_\psi(\theta)]
\right)d\theta^\prime d \theta\, .
\end{equation}
For this we  write 
\begin{equation}
 H[y]\, =\, G[y+\phi^\gd(y)]-G[y]\, .
\end{equation}
We have
\begin{equation}
 DH[y]\, =\, DG[y+\phi^\gd(y)]-DG[y]+DG[y+\phi^\gd(y)]D\phi^\gd(y)
\end{equation}
and thus, using the estimates of theorem \ref{th:M} and the fact that $DG$ is uniformly continuous on a neighborhood of $M$, we get that
\begin{equation}
  \sup_{\psi\in \bbS}  \vert T^\prime_2(\psi)\vert \, \le \, l(\gd) \, .
\end{equation}
with $l(\gd)=o(\gd)$ when $\gd\rightarrow 0$.

 \qed

\section{On the persistence of normally hyperbolic manifolds}
\label{sec:SY} 

In this section we prove theorem \ref{th:M}. The proof in a more general case can be found in \cite{cf:SellYou} but we pay more attention
on the relation between the various small parameters
that enter the proof. We first give a lemma which defines a parametrisation in a neighbourhood of $M$
using the scalar structure given by the operators $L_q$. The proof of this lemma is in \cite[p. 501]{cf:SellYou}.

\begin{lemma}
\label{lem:parametrisation}
There exists a $\gs>0$ such that for all $p$ in the neighborhood 
\begin{equation}
\label{eq:Nsigma}
N_\gs\, :=\, 
\cup_{q\in M} B_{L^2}(q,\sigma)\, ,
\end{equation} 
 of M there is one
and only one $q=v(p) \in M$ such that $(p-q,q')_{-1,1/q}=0$. Furthermore the mapping $p \mapsto v(p)$ is in
$C^\infty(L^2_1,L^2_1)$, and
\begin{equation}
 Dv(p)\, =\, P^o_{v(p)}\, .
\end{equation}
Moreover, the analogous statement holds if 
$N_\gs$ is replaced by $\cup_{q\in M} B_{H_{-1}}(q,\sigma)$ and this time $p \mapsto v(p)$
is  in $C^\infty(\tilde H_{-1},\tilde H_{-1})$.
\end{lemma}
\medskip

For the proof we look for 
 conditions on $\gd$ in order to get a  manifold, which is invariant for for \eqref{eq:main}, at distance $\gep$ from $M$: the condition in the end is going to be that $\gd$ needs to be smaller than a suitable constant times $\gep$
 (and $\gep$ sufficiently small too), so that the invariant manifold is in a neighborhood 
 of order $\gd$ of $M$. To simplify notations,
we will write $F[u]=\partial_\theta(u J*u)$, and \eqref{eq:main} becomes:
\begin{equation}
\label{eq:main3}
\partial_t p_t\, =\, \frac12 \partial^2_\theta p_t - F[p_t]+\gd G[p_t]
\,  .
\end{equation}
We will consider solutions with initial condition $p_0$ satisfying $\Vert p_0-v(p_0) \Vert_2 \leq \gep$. We need asumptions on $\gep$ and $\gd$ such that
the solution stays in $N_\gs$ for a sufficiently long time. If $q$ is in $M$, $w_t:=p_t-q$ satisfies
\begin{equation}
 w_t\, =\, e^{-tL_q}w_0+\int_0^te^{-(t-s)L_q}(F[w_s]+\gd G[q+w_s])\dd s \, ,  
\end{equation}
and we get
\begin{equation}
\Vert w_t\Vert_2\, \leq\, \Vert w_0\Vert_2+\int_0^t\Vert e^{-(t-s)L_{q}}\Vert_{\mathcal{L}(H_{-1},L_2)}(\Vert F[w_s]\Vert_{H_{-1}}+\gd\Vert G[q+w_s]\Vert_{H_{-1}})\dd s\, .
\end{equation}
Define
\begin{equation}
 t_0\,=\,\sup \{ t\geq 0: \, \Vert w_s \Vert_2\leq \gs \text{ for every } s\leq t\}\, .
\end{equation}
Because of the continuity of $w_t$, $t_0>0$ if we suppose $\gep <\gs$. If $t\leq t_0$, using the spectral properties of $L_q$ and the regularity of $F$ and $G$, we get the bounds
\begin{equation}
\label{bound Lq}
 \Vert e^{-(t-s)L_{q}}\Vert_{\mathcal{L}(H_{-1},L_2)}\, \leq\,  C_L(1+(t-s)^{-1/2})\, ,
 \end{equation}
 \begin{equation}
\label{bound G 0}
 \Vert G[q+w_s]\Vert_{H_{-1}}\, \leq\, C_G(1+\Vert w_s\Vert_2)\, ,
 \end{equation}
 and 
 \begin{equation}
\label{bound F 0}
 \Vert F[w_s]\Vert_{H_{-1}}\, \leq \, C_F\Vert w_s\Vert^2_2\, ,
\end{equation}
and thus for all $t_1<t_0$
\begin{equation}
\label{eq:bound wt}
 \Vert w_{t_1} \Vert_2\, \leq\, (\gep + C_G C_L (t_1 + 2\sqrt{t_1})\gd) +C_L(C_F \sigma + C_G \gd)\int_0^{t_1} \left(1+\frac{1}{\sqrt{t_1-s}}\right) \Vert w_s \Vert_2 \dd s\, .
\end{equation}
We need the following lemma, that is a version of the Gronwall-Henry inequality

\medskip 

\begin{lemma}
\label{lem:gron}
 Let $t\mapsto y_t$ be a non-negative and continuous function on $[0,T)$ satisfying for all $t\in[0,T)$ 
\begin{equation}
 y_t\, \leq\, \eta_0 +\eta_1 \int_0^t \left(1+\frac{1}{\sqrt{t-s}}\right)y_s \dd s\, .
\end{equation}
 Then for all $t \in [0,T)$
\begin{equation}
 y_t \, \leq\,  2 \eta_0 e^{\ga t}\, ,
\end{equation}
with $\ga=2 \eta_1 + 4 \eta_1^2 \left(\gG\left(\frac12\right)\right)^2$ where $\gG(r)=\int_0^\infty x^{r-1} e^{-x}\dd x$.
\end{lemma}

\medskip

\noindent
\textit{Proof of lemma \ref{lem:gron}}
We consider the time
\begin{equation}
 t^* \, =\, \sup\{t\geq 0, y_s \leq 2 \eta_0 e^{\ga s}\text{ for all }s\leq t\}\, .
\end{equation}
We have to show that $t^*=T$. But if $t^*< T$, then
\begin{multline}
 y_{t^*} \, \leq\, \eta_0\left(1 +2\eta_1\int_0^{t^*}\left(1+\frac{1}{\sqrt{t^*-s}}\right)e^{\ga s}ds\right)\\
          \leq\, \eta_0\left(1+\frac{2\eta_1}{\ga}[e^{\ga t^*}-1]+\frac{2 \eta_1}{\sqrt{\ga}}\gG\left(\frac12\right)e^{\ga t^*}\right)\, 
         < \, 2 \eta_0 e^{\ga t^*}\, ,
\end{multline}
which contradicts $t^*< T$ since $y_\cdot$ is continuous.
\qed

\medskip
\noindent
Using Lemma \ref{lem:gron} and \eqref{eq:bound wt} we get :
\begin{equation}
\label{bound wt}
 \Vert w_t \Vert_2 \, \leq\,  C(t_1)(\gd+\gep)\, ,
\end{equation}
where
\begin{equation}
 C(t_1)\, =\, \max(1,C_G C_L (t_1+2\sqrt{t_1}))e^{\left(2 \eta(\gs,\gd) + 4 \pi\eta(\gs,\gd)^2 \right)t_1}\, ,
 \end{equation}
 \begin{equation}
\eta(\gs,\gd)\, =\, C_L(C_F \sigma + C_G \gd)\, .
\end{equation}
For $T>0$, if we choose $\gep$ and $\gd$ such that $C(2T)(\gep+\gd)\leq \sigma$, then $p_t$ lies in
$N_\gs$ for $t\in[0;2T]$. Take now $T$ such that
\begin{align}
 \label{hyp T}
&C_{P^s} e^{-\lambda_1 T/2}\, \leq\,  \frac{1}{16}\, ,
\\
 \label{hyp T 2}
&e^{\lambda_1 T/2}\, \geq\, 4 C_L\, ,
\end{align}
where we recall that $\gl_1$ is the spectral gap of $L_q$ and we set
\begin{equation}
\label{CPs}
 C_{P^s}\, =\,\max_{q\in M} \Vert P^s_{q}\Vert_{\cL(L^2_0,L^2_0)}\, ,
\end{equation}
and $P^s_q$ is a compact notation for $P^s (q)$ (defined just below \eqref{eq:shyp1})
and it is the orthogonal projection of the range of $L_q$ (the scalar product is the one of $H_{-1,1/q}$).
Define also
\begin{equation}
 \label{eq:defalso6}
C_1\, =\, C(2T), \ \ \ 
C_2\, =\, e^{\lambda_1 T/2} \ \text{ and } \ 
\gep_0\, =\, \frac{\gs}{2C_1}\, .
\end{equation}
For now we will take $\max\{\gep,\gd\}\leq \gep_0$, so that $p_t \in N_\gs$
for $t \le 2T$. We will use the following notations:
\begin{equation}
p_i\, :=\, p(t,p_{i0})\, ,
\end{equation}
is the solution of \eqref{eq:main3}
and
\begin{equation}
v_i\, :=\, v_i(t,p_{i0})\,:=\, v(p_i)\, ,
\end{equation}
is given by Lemma
  \ref{lem:parametrisation}. Moreover we set
  \begin{equation}
n_i\, =\, p_i-v_i\, , \ \ \
\gD p\, :=\, p_1-p_2 \, ,\ \ \
\gD v\, :=\, v_1-v_2\, , \ \ \
\gD n\, :=\, n_1-n_2\, .
\end{equation}
In the following lemma we compare the quantites we have just introduced with the initial conditions. It corresponds to Lemma 74.7 (page 507) in \cite{cf:SellYou}.
We remark that is in this lemma $\gep$ and $\gd$ play the same role and we stress that these
are just preliminary estimates: some of them are going to be refined later on.

\begin{lemma}
\label{lem:bound traj}
 For all $\ga>0$, there exist $C_0=C_0(T)$ and $\gep_1\leq \gep_0$ such that if $\gep \leq \gep_1$ and $\delta\leq \gep_1$ we have the following properties:
\begin{enumerate}
\item if $\Vert p_0-v_0 \Vert_2 \leq \gep$ then for all $t\in[0,2T]$
\begin{equation}
\label{eq:3in1-1}
\max\left(
\Vert p(t,p_0)-v_0\Vert_2 \, ,\,  
\Vert v(t,p_0)-v_0\Vert_2 \, ,\,   \frac{1}{2}\Vert n(t,p_0)\Vert_2 \right)\,  \leq\,  C_0(\gep+\delta) \, ;
\end{equation}
\item if $\Vert p_{i0}-v_{i0}\Vert_2\leq \gep$ and $\Vert \Delta v(0)\Vert_2\leq \ga \gep$, then for all $t\in[0,2T]$
\begin{equation}
\max\left(
\Vert\Delta p(t)\Vert_2  \, , \, 
\Vert\Delta v(t)\Vert_2   \, , \,
\Vert\Delta n(t)\Vert_2 \right)\, 
\leq \,  C_2\Vert \Delta p(0)\Vert_2 \, ,
\end{equation}
with $C_2$ given in \eqref{eq:defalso6};
\item if $\Vert p_{i0}-v_{i0}\Vert_2\leq \gep$ and $\Vert\Delta p(0)\Vert_2\leq 2\Vert \Delta v(0)\Vert _2$, then for all $t\in[0,2T]$
\begin{equation}
\frac{1}{2}\Vert \Delta v(0)\Vert_2\, \leq\, \Vert \Delta v(t)\Vert_2 \, \leq\, \frac{3}{2}\Vert \Delta v(0)\Vert_2\, .
\end{equation}
\end{enumerate}
\end{lemma}

\medskip

\noindent
\textit{Proof of Lemma \ref{lem:bound traj}}
For what concerns part (1)
note that the first of the three inequalities in \eqref{eq:3in1-1} is given above (see \eqref{bound wt} with $t_0=2T$). The other inequalities  come from the fact that the mapping $q\mapsto v(q)$ of 
Lemma~\ref{lem:parametrisation} is Lipschitz, taking, if necessary, a bigger value for $C_0$. 
\medskip

For part (2)notice that, since $v_{20}\in M$, we can write the evolution in mild form around $v_{20}$, that is
\begin{multline}
 \Delta p(t) \, =\,  e^{-t L_{v_{20}}}\Delta p(0)\\
 +\int_0^te^{-(t-s)L_{v_{20}}}(F[p_1(s)-v_{20}]-F[p_2(s)-v_{20}]+\gd(G[p_1(s)]-G[p_2(s)]))\dd s\, ,
\end{multline}
and thus
\begin{equation}
\label{bound Delta p}
 \Vert \Delta p(t)\Vert_2\, \leq\, C_L\Vert \Delta p(0)\Vert_2 +C_L\Big(C_F(\ga\gep+C_0(\gep+\gd)+C_G\gd\Big)\int_0^t\left(1+\dfrac{1}{\sqrt{t-s}}\right)\Vert \Delta p(s)\Vert_2 \dd s\, .
\end{equation}
Here we used the preceding point, \eqref{bound Lq} and the bounds
\begin{equation}
\label{bound lipsch G}
  \Vert G[p_1(s)]-G[p_2(s)]\Vert_{-1} \, \leq \, C_G \Vert p_1(s)-p_2(s)\Vert_2 \, , 
  \end{equation}
 \begin{equation}
\label{bound lipsch F}
  \Vert F[p_1(s)]-F[p_2(s)]\Vert_{-1} \, \leq\,  C_F (\ga \gep + C_0 (\gep+\gd))\Vert p_1(s)-p_2(s)\Vert_2 \, ,
\end{equation}
\eqref{bound lipsch F} is obtained by applying the mean value inequality to $F$ and $DF$ and using the fact that $DF(0)=0$:  the constants $C_G$ and $C_F$ have a larger value than in \eqref{bound G 0} and \eqref{bound F 0}. 
%The constants $C_G$ and $C_F$ have a larger value than in \eqref{bound G 0} and \eqref{bound F 0} since $\Vert p_{1s}-p_{2s}\Vert_2$ may be larger than $\sigma$.
Applying Lemma \ref{lem:gron} to \eqref{bound Delta p}, we obtain
\begin{equation}
\Vert \Delta p(t) \Vert_2\,  \leq\, 2 C_L e^{\left(2 \eta_1(\gep,\gd) + 4 \pi\eta_1(\gep,\gd)^2 \right)2T}  \Vert\Delta p(0)\Vert_2
\end{equation}
with
\begin{equation}
 \eta_1(\gep,\gd)\, =\, C_L\Big(C_F(\ga\gep+2C_0(\gep+\gd))+C_G\gd\Big)\, ,
\end{equation}
Choose $\gep_1\leq \gep_0$ such that (it is possible because of \eqref{hyp T 2})
\begin{equation}
 2 C_L e^{\left(2 \eta_1(\gep_1,\gep_1) + 4\pi \eta_1(\gep_1,\gep_1)^2 \right)2T}\, \leq\, e^{\lambda_1 T/2} \, .
\end{equation}
The two other points come directly from the Lipschitz property of the mapping $q\mapsto v(q)$ taking, if necessary, a smaller value for $\gep_1$.
\medskip

For part (3) we prove first that for all $r>0$, there exists $\gep_2(r)$ such that for all $\gep\leq \gep_2(r)$ and $\delta\leq \gep_2(r)$ we have for all $t\in[0,2T]$
\begin{equation}\frac{1}{2}\leq \frac{\Vert\Delta v(t)\Vert_2}{\Vert\Delta v(0)\Vert_2}\leq \frac{3}{2}\end{equation}
if $\Vert\Delta v(0)\Vert_2\geq r$. In fact, in this case, using Lemma \ref{lem:parametrisation} :
\begin{align}
\left|\dfrac{\Vert \Delta v(t)\Vert_2-\Vert \Delta v(0)\Vert_2}{\Vert \Delta v(0)\Vert _2}\right|  &\, \leq\, \dfrac{|\ \Vert \Delta v(t)\Vert_2-\Vert \Delta v(0)\Vert_2\ |}{r}\notag\\
                  &\, \leq\, \dfrac{\Vert \Delta v(t)-\Delta v(0)\Vert_2}{r}\\
                  &\, \leq\,  \dfrac{\Vert v_1(t)-v_1(0)\Vert_2+\Vert v_2(t)-v_2(0)\Vert_2}{r}\notag\\
                  &\, \leq\,  \dfrac{2 C_0 (\delta +\gep)}{r}\notag \, .
\end{align}
We can choose $\gep_2(r)=\min(\gep_1,r/8C_0)$. Now it is sufficient to prove that, for $\Vert \Delta v(0)\Vert_2\leq r_0$ with a certain $r_0$, 
\begin{equation}
\Vert \Delta v(t)- \Delta v(0)\Vert_2\, \leq\, \frac{1}{2}\Vert \Delta v(0)\Vert_2
\end{equation}
for all $t\in[0,2T]$. Suppose that $\Vert \Delta v(0)\Vert_2 \leq r$ with $r\leq \ga$. We use the following decomposition 
\begin{align}
\Delta v(t) - \Delta v(0)\, =\, &\Delta v(t)-P^o_{v_2}\Delta p(t)-\Delta v(0)+P^o_{v_{20}}\Delta p(0)\\ \notag
                          &+(P^o_{v_2}-P^o_{v_{20}})\Delta p(t)+P^o_{v_{20}}(\Delta p(t) -\Delta p(0))\, .
\end{align}
From Lemma \ref{lem:parametrisation} , part (2) and the hypothesis $\Vert \Delta p(0)\Vert_2\leq 2\Vert \Delta v(0)\Vert_2$, we get
\begin{align}
\notag
\Vert \Delta v(t)-P^o_{v_2}\Delta p(t)\Vert_2 &\, =\, \Vert v(p_1)-v(p_2)-Dv_{v_2}(p_1-p_2)\Vert_2 \\
                                              &\, \leq\, 2C_v (r+2C_0(\delta+\gep))\Vert \Delta v(0)\Vert_2\\
\notag
\Vert \Delta v(0)-P^o_{v_{20}}\Delta p(0)\Vert_2&\, =\, \Vert v(p_{10})-v(p_{20})-Dv_{v_{20}}(p_{10}-p_{20})\Vert_2 \\
\label{bound taylor v}
                                                &\, \leq\, 4C_v r \Vert \Delta v(0)\Vert_2
\end{align}
where $C_v$ is the maximum of the second derivate of $q\mapsto v(q)$ in $N_\gs$. Since $P^o$ is $C^1$ on $M$, there exits $L_{P^o}$ such that (By using part (1) and the hypothesis and defining $L_{P^o}$ the maximum of the norme of $DP$ on $M$, we get
\begin{equation}
 \Vert (P^o_{v_2}-P^o_{v_{20}})\Delta p(t)\Vert_2\leq L_{P^o} C_0(\delta+\gep) \Vert \Delta v(0)\Vert_2
\end{equation}
For the last term, we write
\begin{align}
\notag
\Delta p(t) -\Delta p(0)\, =\, (e^{-t L_{v_{20}}}-I)\Delta p(0)+\int_0^te^{-(t-s)L_{v_{20}}}(&F[p_{1s}-v_{20}]-F[p_{2s}-v_{20}]\\
                                                                                     &+G[p_{1s}]-G[p_{2s}])ds\, .
\end{align}
Notice that $P^o_{v_{20}}(e^{-t L_{v_{20}}}-I)\Delta p(0)=0$. From \eqref{bound Lq}, \eqref{bound lipsch F} and \eqref{bound lipsch G} it comes
\begin{equation}
\label{bound integral term}
\Vert P^o_{v_{20}} ( \Delta p(t)-\Delta p(0) )\Vert_2\, \leq\, 4 C_{P^o}  C_L C_2 \Big(C_F(r+2C_0(\delta+\gep))+C_G\delta\Big)(T+\sqrt{2T}) \Vert\Delta v(0)\Vert_2
\end{equation}
where $C_{P^o}$ is the maximum of the norms $\Vert P^o_q\Vert_{\cL(L^2_0,L^2_0)}$ for $q\in M$.
In conclusion there exists a constant $C_3$ such that for all $t\in [0,2T]$
\begin{equation}
\Vert \Delta v(t)- \Delta v(0)\Vert_2\, \leq\, C_3(r+\gep+\delta)\Vert\Delta v(0)\Vert_2\, .
\end{equation}
To end the proof choose $r=r_0:=\min\left(\ga,\frac{C_3}{3}\right)$ and reduce if nececessary the value of $\gep_1$ to have $\gep_1 \leq \min(\gep_2(r_0),r_0)$.

\qed

\medskip

We now  move to the main body of the proof which is based on introducing a family of transformations
of the manifold $M$ by using the full dynamics and we aim at identifying the transformation that maps $M$
to the manifold that is stable for the full dynamics and this is achieved by 
 applying the Banach fixed point Theorem in a relevant space of functions.
 
Define the set $C(M,L^2_0)$ of continuous functions from $M$ to $L^2_0$ provided with the norm
\begin{equation}
 \Vert f \Vert_\infty\, =\, \sup \{ \Vert f(v) \Vert_{L^2}\ ,\ v\in M \}
\end{equation}
and consider the subset $\cF (\gep,l)$ of $C(M,L^2_0)$ of functions $f$ satisfying :
\begin{enumerate}
\item $\Vert f \Vert_\infty \leq \gep$
\item $f$ is Lipschitz on $M$ with Lipschitz constant $l\leq 1$
\item $(f(q),q')_{-1,1/q}=0$ for all $q$ in $M$
\end{enumerate}
Notice that $\cF (\gep,l)$ is a complete subset of $C(M,L^2_0)$. We will now define
a set of mappings $\left\{ X_t \right\}_{t\in [T,2T]}:\cF (\gep,1)\mapsto C(M,L^2_0)$ and show that 
\begin{enumerate}
 \item for all $\tau\in[T,2T]$
    \begin{equation}
      \label{prop Xt 1}
      X_\tau \left(\cF (\gep,1)\right)\, \subset\, \cF \left(\gep,\frac{1}{4C_2}\right) 
    \end{equation}
 (recall that $C_2=e^{\lambda_1 T/2}$ and thus $\frac{1}{4C_2}\leq 1$)
 \item $X_T$ is a contraction on $\cF (\gep,1)$:

    \begin{equation}
       \label{prop Xt 2}
       \Vert X_T(f_1)-X_T(f_2) \Vert_\infty\, \leq\, \frac12 \Vert f_1-f_2\Vert_\infty 
    \end{equation}
       for all $f_1,f_2\in \cF (\gep,1)$.
\end{enumerate}
Notice that the third point of \eqref{lem:bound traj} and an argument of connexion ( see \cite{cf:SellYou} page 513 ) show that for all
$f\in \cF (\gep,1)$ and $t\in [0,2T]$, the mapping $q\mapsto g_{t,f}(q):=v(t,f(q))$ is a bijection of $M$. So we can define the mappings
\begin{equation}
\begin{array}{cccc}
 X_\tau(f)(u): & M & \rightarrow & L^2_0 \\
               & u &  \mapsto    & n\left(\tau,(i_d+f)\circ g_{\tau,f}^{-1}(u)\right)

\end{array}
\end{equation}
It is easy to see that for all $\tau \in [T,2T]$ and $f\in \cF (\gep,1)$, $X_\tau(f)$ is the unique
mapping satisfying for all $q\in M$
\begin{equation}
\label{eq:prop Xtau}
 X_\tau(f)(v(\tau,p_0))\, =\, n(\tau,p_0)=p(\tau,p_0)-v(\tau,p_0)=P^s_{v(\tau,p_0)}(p(\tau,p_0)-v(\tau,p_0))
\end{equation}
where $p_0=q+f(q)$. We can see $X_t(f)$ as the {\sl distance} (in the sense of \eqref{lem:parametrisation}) of
the trajectory $p_t$ from $M$, starting at the time $0$ at a {\sl distance} $f$ from $M$.

\medskip

In the following, we will first prove that \eqref{prop Xt 1} and \eqref{prop Xt 2} imply that there exists an  invariant manifold $M_\gep$ for \eqref{eq:main3} at distance
$\gep$ of $M$. Then we will prove \eqref{prop Xt 1} and \eqref{prop Xt 2} in three lemmas, paying attention on the relations between the different parameters. 

\medskip

\noindent
Suppose that the mappings $X_\tau$ satisfy \eqref{prop Xt 1} for $\tau\in,[T,2T]$ and that $X_T$ satisfies \eqref{prop Xt 2}. Then $X_T$ has a unique
fixed point in $\cF (\gep,1)$, which will be noted $f_0$. Define $\phi^\gep=id+f_0$ on $M$ and $M_\gep=\phi_0(M)$.
Since $f_0$ is a fixed point of $X_T$, if $p_0\in M_0$, then $p_{kT} \in M_0$ for all $k\in \bbN$.
Then to prove that $M_\gep$ is an invariant manifold of \eqref{eq:main3}, it is sufficient to prove that
for all $t\in(0,T)$, the functions $f_t$ defined by $f_t=X_t(f_0)$ are equal to $f_0$. Using the property of
semi-group and $X_T(f_0)=f_0$ it is easy to see that $f_t=X_{T+t}(f_0)$, and thus \eqref{prop Xt 1} implies
that $f_t \in \cF (\gep,1)$. But the same arguments show that $f_t$ is a fixed point of $X_T$ for all
$t\in (0,T)$. In conclusion, $M_\gep$ is invariant for \eqref{eq:main3}.

\medskip

Now we prove \eqref{prop Xt 1} and \eqref{prop Xt 2} in the three following lemmas, which correspond to Lemmas 74.8, 74.9 and 74.10 in \cite{cf:SellYou}.

\begin{lemma}
 \label{lem:bound Xtau}
There exists a $\gep_3\leq \gep_1$ such that if $\gep\leq \gep_3$, there exists a $\delta_3(\gep)$ of the form $\min(C\gep,\gep_3)$ such that if $\delta\leq \delta_3(\gep)$,
we have for all $\tau\in[T,2T]$ and $f\in \cF(\gep,1)$
\begin{equation}\Vert X_\tau(f)\Vert_\infty\leq \gep\end{equation}
\end{lemma}

\medskip

\noindent
\textit{Proof}
Let $v_0\in M$, $p_0 =v_0+f(v_0)$. We write (see \eqref{eq:prop Xtau})
\begin{equation}
\label{eq:lem bound Xtau}
 X_\tau(f)(v(\tau))\, =\, P^s_{v(\tau)}(p(\tau)-v_0)-P^s_{v(\tau)}(v(\tau)-v_0)\, .
\end{equation}
The first term can be written as
\begin{equation}
\label{eq:first term lem bound Xtau}
 P^s_{v(\tau)}(p(\tau)-v_0)\, =\, P^s_{v(\tau)}\left(e^{-\tau L_{v_0}}(p_0-v_0)+\int_0^\tau F[p(s)-v_0]+G[p(s)]ds\right)\, .
\end{equation}
Using the spectral gap, we bound the linear term
\begin{equation}
 \Vert P^s_{v(\tau)} e^{-\tau L_{v_0}}(p_0-v_0) \Vert_2\, \leq\, C_{P^s} e^{-\lambda_1 \tau} \gep
\end{equation}
and the remaining term of \eqref{eq:first term lem bound Xtau} can be bounded in the same way as \eqref{bound integral term}. 
Furthemore the second term of \eqref{eq:lem bound Xtau} is quadratic in $\gep$ and $\gd$, using a Taylor argument  as in \eqref{bound taylor v}.
Finaly, we get
\begin{equation}
 \Vert X_\tau(f)(v_\tau) \Vert_2\, \leq\, C_4\Big( (\gd+\gep)^2+\gd)\Big) + C_{P^s} e^{-\lambda_1 \tau} \gep \, .
\end{equation}
We supposed $C_{P^s} e^{-\lambda_1 T}\leq \frac{1}{16}$, thus we can choose
\begin{equation}
 \gep_3 \, =\, \min\left(\gep_1,\frac{1}{12 C_4}\right)\ \text{ and } \ 
 \gd_3(\gep)\, =\, \min\left(\gep_1,\gep,\frac{1}{3 C_4}\gep\right)\, .
\end{equation}

\qed

\begin{lemma}\label{lem:lipsch X(f)}
There exists $\gep_4\leq \gep_3$ such that if $\gep\leq \gep_4$, there exists a $\gd_4(\gep)$ of the form $\min(C\gep,\gep_4)$ such that if $\gd\leq \gd_4(\gep)$,
then for all $f\in\mathcal{F}(\gep,1)$ we have $X_\tau(f)\in\mathcal{F}\left(\gep,\frac{1}{4C_2}\right)$
for all $\tau\in[T,2T]$.
\end{lemma}

\medskip

\noindent
\textit{Proof}
It is sufficient to prove that $X_\tau(f)$ is Lipschitz with Lipschitz constant $\frac{1}{4C_2}$ on all $M \cap B_2(q,\rho_0)$ with $\rho_0=8C_2\gep$. Indeed in this case, if $\Vert q_1-q_2 \Vert_2 > \rho_0$, then
\begin{equation}
\label{eq:g794}
\Vert X_\tau(f)(q_1)-X_\tau(f)(q_2)\Vert_2\, \leq\, \frac{2 \gep}{\rho_0}\Vert q_1-q_2\Vert_2\, \leq\, \frac{1}{4 C_2}\Vert q_1-q_2\Vert_2\, .
\end{equation}
Take $u_1,u_2\in M$ such that $\Vert u_1-u_2 \Vert_2 \leq \rho_0$ and $f$ with Lipschitz constant $l\leq 1$. There exists $v_{10},v_{20} \in M$ such that $u_i=v(\tau,p_{i0})$ with
$p_{i0}=v_{i0}+f(v_{i0})$. Our goal is to show that under the hypothesis
\begin{equation}
\dfrac{\Vert X_\tau(f)(u_1)-X_\tau(f)(u_2)\Vert_2}{\Vert u_1-u_2\Vert_2}\, =\,
\dfrac{\Vert X_\tau(f)(v_1(\tau))-X_\tau(f)(v_2(\tau))\Vert_2}{\Vert v_1(\tau)-v_2(\tau)\Vert_2}\, =\, \dfrac{\Vert\Delta n(\tau)\Vert_2}{\Vert\Delta v(\tau)\Vert_2}\leq \dfrac{1}{4C_2}\, .
\end{equation}
We use the decomposition
\begin{align}
\notag
\Delta n(\tau)&\, =\,  e^{-\tau L_{v_{20}}}P^s_{v_{20}}\Delta n(0)+ \Delta n(\tau)-e^{-\tau L_{v_{20}}}P^s_{v_{20}}\Delta n(0)\\
\notag
              &\, =\, e^{-\tau L_{v_{20}}}P^s_{v_{20}}\Delta n(0) +\Delta p(\tau) -\Delta v(\tau) -e^{-\tau L_{v_{20}}}P^s_{v_{20}}\Delta p(0)+e^{-\tau L_{v_{20}}}P^s_{v_{20}}\Delta v(0)\\
\notag
              &\, =\,  e^{-\tau L_{v_{20}}}P^s_{v_{20}}\Delta n(0)+\Delta p(\tau)-P^o_{v_{20}}\Delta p(t)+P^o_{v_{20}}\Delta p(t)-\Delta v(\tau) -e^{-\tau L_{v_{20}}}P^s_{v_{20}}\Delta p(0)\\
\notag
              &\quad  \; +e^{-\tau L_{v_{20}}}P^s_{v_{20}}\Delta v(0)\\
\notag
              &\, =\, \left[e^{-\tau L_{v_{20}}}P^s_{v_{20}}\Delta n(0)\right]+\left[(P^s_{v_2(\tau)}-P^s_{v_{20}})\Delta p(\tau)\right] +\left[P^s_{v_{20}}(\Delta p(\tau)- e^{-\tau L_{v_{20}}}\Delta p(0))\right]\\
\notag
              & \quad  \; +\left[e^{-\tau L_{v_{20}}}P^s_{v_{20}}\Delta v(0)\right]+\left[P^o_{v_2(\tau)}\Delta p(\tau)-\Delta v(\tau)\right]\, .
\end{align}
We bound the first term using the spectral gap of $L_{v_{20}}$, the second term using the smoothness of $P^s$ and Lemma \ref{lem:bound traj}, and the third term in a
similar way as \eqref{bound integral term}. We use a Taylor decomposition for the two last terms, as in \eqref{bound taylor v}. Then we get (recall \eqref{CPs})
\begin{equation}
\Vert\Delta n(\tau)\Vert_2\, \leq\,  \Big( C_{P^s} e^{-\lambda_1 T}l+C_5(\rho_0+\delta+\gep)\Big)\Vert\Delta v(0)\Vert_2\, .
\end{equation}
Since
$f$ is Lipschitz with Lipschitz constant $l\leq 1$ we have  $\Vert \Delta p(0)\Vert_2 \leq 2\Vert \Delta v(0) \Vert_2$. Then using the part (3) of
Lemma \ref{lem:bound traj} we deduce 
\begin{equation}
 \Vert \Delta v(0)\Vert_2\, \leq\, 2\Vert\Delta v(\tau) \Vert_2 \, .
\end{equation}
Furthemore we have chosen $T$ such that $C_{P^s} e^{-\lambda_1 T/2}\leq \frac{1}{16}$, and thus $C_{P^s} e^{-\lambda_1 T} \leq \frac{1}{16C_2}$ (recall that $C_2=e^{\lambda_1/2}$). We obtain
\begin{equation}
\label{ineq:lipsch arg}
\dfrac{\Vert\Delta n(\tau)\Vert_2}{\Vert\Delta v(\tau)\Vert_2}\, \leq\, \frac{1}{8C_2}l+2C_5((1+8C_2)\gep+\gd)\, .
\end{equation}
Finally choose
\begin{equation}
 \gep_4\, =\, \min\left(\gep_3,\dfrac{1}{32C_2 C_5 (1+4C_2)}\right)\ \text{ and } \ 
 \gd_4(\gep)\, =\,\min\left(\gep_4,\gd_3(\gep)\right)\, ,
\end{equation}
and the proof is complete.
\qed

\begin{lemma}
\label{lem:contraction XT}
There exists $\gep_5\leq \gep_4$ such that if $\gep\leq \gep_5$, there exists a $\gd_5(\gep)$ of the form $\min(C\gep,\gep_5)$ such that
for all $f_i\in \mathcal{F}\left(\gep,\frac{1}{4C_2}\right)$:
\begin{equation}
\Vert X_T(f_1)-X_T(f_2)\Vert_\infty\, \leq\,  \frac{1}{2} \Vert f_1-f_2 \Vert_\infty\, .
\end{equation}
\end{lemma}

\medskip

\noindent
\textit{Proof}
This time take $v_{10}=v_{20}=v_0$ and $p_{i0}=v_0+f_i(v_0)$. With the same decomposition as in Lemma \ref{lem:lipsch X(f)} (with fewer terms, since
$v_{10}=v_{20}$) we get
\begin{equation}
 \Vert \Delta n(T) \Vert_2 \, 
 \leq\,  \Big( C_{P^s} e^{-\lambda_1 T} + C_6 (\gd+\gep)\Big) \Vert \Delta p(0) \Vert_2\, .
\end{equation}
We choose
\begin{equation}
 \gep_5 \, =\, \min\left( \gep_4, \dfrac{1}{16 C_6}\right)\ \text{ and } \
 \gd_5(\gep)\, =\, \min\left(\gep_5, \gd_4(\gep)\right)\, ,
\end{equation}
and in this case we get 
\begin{equation}
 \Vert \Delta n(T) \Vert_2 \, \leq \, \frac14 \Vert f_1-f_2 \Vert_\infty\, .
\end{equation}
Now notice that
\begin{equation}
\Vert(X_T(f_1)-X_T(f_2))(v_2(T))\Vert_2\, \leq\,  \Vert\Delta n(T)\Vert_2+\Vert X_T(f_1)(v_1(T))-X_T(f_1)(v_2(T))\Vert_2\, ,
\end{equation}
and since $X_T(f_1)$ is Lipschitz with Lipschitz constant $\frac{1}{4C_2}$, we get, using Lemma \ref{lem:parametrisation}
\begin{equation}
 \Vert X_T(f_1)(v_1(T))-X_T(f_1)(v_2(T))\Vert_2\, \leq\,  \dfrac{1}{4 C_2} \Vert \Delta v(T) \Vert_2 \,
 \leq\,  \dfrac14 \Vert f_1 -f_2 \Vert_\infty\, .
\end{equation}
\qed

\medskip

\noindent
{\it Proof of Theorem~\ref{th:M}.}
In these three lemmas, we see that if $\gep$ is small enough, we can take $\gd$ proportional to $\gep$, thus adding a perturbation
of type $\gd G[p_t]$ to \eqref{eq:Kuram} creates an invariant manifold $M_\gd$ situated at a distance $O(\gd)$
 from
$M$. It is proven in \cite[(theorem 74.15, p.~531)]{cf:SellYou}  that the manifold $M_\gd$ is $C^1$ in $L^2_1$ and normally hyperbolic. Remark furthermore that
$\left(\phi^\gd\right)^{-1}(p)=v(p)$ for all $p\in M_\gd$. So to prove that $\phi^\gd$ is $C^1$, it suffices to prove that $v$ satisfies the hypothesis of
the local inverse theorem between manifolds, that is $Dv$ is a bijection between the tangent spaces of de two manifolds. Since the manifold is of dimension one,
this property is implied by the lipschitz property  of $\phi^\gd$. Furthermore we can estimate the differential of $\phi^\gd$ : \eqref{ineq:lipsch arg} for $\phi^\gd$ gives
an inequality for the local Lipschitz constant $l^\gd$ of $\phi_\gd$ on all neighborhoods $M\cup B_2(q,\rho_0)$ 
($\rho_0$ is introduced right before \eqref{eq:g794}):
\begin{equation}
 l^\gd \, \leq\,  \dfrac{1}{8C_2} l^\gd+ C_7 \gd\, ,
\end{equation}
and we get that for a $C_8>0$
\begin{equation}
 l^\gd\, \leq\,  C_8 \gd\, ,
\end{equation}
which yields the bound we claim on the differential of $\phi_\gd$.
\qed

\appendix

\section{On a norm equivalence}
\label{app:normeq}
The goal is to prove that the norms $\Vert \cdot \Vert_{H_1}$ and $\Vert \cdot \Vert_{V_q^2}$ are equivalent, with
\begin{equation}
 -L_qu\, :=\, \frac12 u'' -[uJ*q+qJ*u]'
\end{equation}
 and
\begin{equation}
 \Vert u \Vert_{V_q^2}\, :=\, \Vert (C+L_q)u \Vert_{-1,1/q}
\end{equation}
with $C>0$. Remark that by changing the constant $C$ we get an equivalent norm.
Since the norms $\Vert \cdot \Vert_{-1,1/q}$ and $\Vert \cdot \Vert_{-1}$ are equivalent, we will study
$\Vert (C+L_q)u \Vert_{-1}$. We write 
\begin{equation}
 u(\theta)\, =\, \sum a_n e^{i n \theta}\, .
\end{equation}
$J*q$ is of the type $\ga e^{i\theta}-\ga e^{-i\theta} $, thus we can write
\begin{equation}
 uJ*q(\theta)\, =\, \ga\left( \sum a_n e^{i(n+1)\theta}-\sum a_n e^{i(n-1)\theta}\right)\, .
\end{equation}
Furthermore 
\begin{equation}
 J*u(\theta)\, =\, -\frac{Ka_1}{2i}e^{i\theta}+\frac{Ka_{-1}}{2i}e^{-i\theta}\, .
\end{equation}
So if we denote
\begin{equation}
 q(\theta)\, =\, \sum c^q_n e^{i\theta}
\end{equation}
then
\begin{equation}
 qJ*u\, =\, -\frac{Ka_1}{2i}\sum c^q_n e^{(n+1)\theta}+\frac{Ka_{-1}}{2i}\sum c^q_n e^{(n-1)\theta}\, .
\end{equation}
Consequently
\begin{multline}
\label{eq:norm}
 \Vert (C+L_q)u \Vert_{-1}\, =\\
  \sum (1+n^2)^{-1}\left|C a_n +n^2 a_n -i\ga n (a_{n-1}-a_{n+1}) +n\frac{Ka_1}{2}c^q_{n-1}-n\frac{Ka_{-1}}{2}c^q_{n+1}\right|^2\, .
\end{multline}
Suppose now that $u\in H_1$. It is easy to see that there exists $c>0$ such that $\Vert u \Vert_{V^2_q}\leq c\Vert u\Vert_{H_1}$. Thus $\sum n^2|a_n|^2<\infty$ implies that
$\Vert (C+L_q)u \Vert_{-1}<\infty$ and so $H_1\subset V_q^2$. By expanding \eqref{eq:norm}
and using Cauchy-Schwartz inequality we get
\begin{equation}
\Vert (C+L_q)u\Vert_{-1}\, \geq\, \sum (1+n^2)^{-1}\left(C^2+ n^4+2C n^2 -\ga_1  n^3-\ga_2 C n \right)|a_n|^2
\end{equation}
where $\ga_1,\ga_2\geq 0$ do not depend on $u$. It is clear that for $C$ big enough ( depending on $\ga_1$ and $\ga_2$ ) we have
\begin{align}
 &\frac{C^2}{2}+n^4-\ga_1 n^3\, \geq\, \frac12 n^4\\
 &\frac{C^2}{2}+2Cn^2-\ga_2 Cn\, \geq\, 0
\end{align}
and thus $\Vert (C+L_q)u\Vert_{-1}\geq \frac14\Vert u\Vert_{H_1}$. We have shown that there exist $c>0$ such that for all $u\in H_1$,
\begin{equation}
 c^{-1}\Vert u\Vert_{V_q^2}\, \leq\, \Vert u\Vert_{H_1} \leq c \Vert u\Vert_{V_q^2}\, .
\end{equation}
But $H_1$ is dense in $V_q^2$ (consider the finite sums of fourier series). If $v\in V_q^2$, there exists a sequence 
$v_n$ in $H_1$ such that $v_n\rightarrow v$ for the $V_q^2$ norm. Then $v_n$ is a Cauchy sequence for the $H_1$ norm, and
since $H_1$ is complete, $v\in H_1$. In conclusion $V_q^2$ and $H_1$ have the same elements.

\medskip
\begin{rem}
By  replacing $(1+n^2)^{-1}$ by $(1+n^2)^{k}$, we can prove in the same way  that $\Vert (C+L_q)u\Vert_{H_k}$ is equivalent to $\Vert u \Vert_{H_{k+2}}$. Thus  $\Vert u \Vert_{V^{n}_q}=\Vert (1+L_q)^{n/2}u\Vert_{-1,1/q}$
is equivalent to $\Vert u \Vert_{H_{n-1}}$.
\end{rem}

\section*{Acknowledgements}
 G.~G. and K.~P. acknowledge the support of the ANR grant ManDy.
G.~G.  acknowledges also the support of ANR grant SHEPI.

\end{document}